\documentclass[10pt]{article}  
\usepackage[margin=1in]{geometry}
\usepackage{graphicx}
\usepackage{url}
\usepackage[utf8]{inputenc}
\usepackage[english]{babel}
 \usepackage{float}
\usepackage{amsthm}
\newtheorem*{remark}{Remark}
\title{A hybrid discrete-continuum approach to model Turing pattern formation \\ {\small Dedicated to the memory of Federica Bubba}}
\author{%
 Fiona R Macfarlane$^{1}$, 
 Mark AJ Chaplain$^{1}$
 and
 Tommaso Lorenzi$^{2}$
}
\date{$^{1}$ School of Mathematics and Statistics, University of St Andrews, UK;\\
$^{2}$ Department of Mathematical Sciences ``G. L. Lagrange'', Dipartimento di Eccellenza 2018-2022, Politecnico di Torino, 10129 Torino, Italy;}

\usepackage{latexsym}
\usepackage{amssymb,amsbsy,amsmath,amsfonts,amssymb,amscd,amsfonts,mathrsfs}
\usepackage{graphicx}
\usepackage{color} 

\usepackage{cite}
\usepackage{txfonts}
\usepackage{mathtools}
\DeclarePairedDelimiter{\ceil}{\lceil}{\rceil}
\newcommand{\beq}{\begin{equation}}
\newcommand{\eeq}{\end{equation}}
\newcommand{\beqa}{\begin{eqnarray}}
\newcommand{\eeqa}{\end{eqnarray}}

\numberwithin{equation}{section}

\begin{document}
\maketitle

\begin{abstract}
Since its introduction in 1952, with a further refinement in 1972 by Gierer and Meinhardt, Turing's (pre-)pattern theory (the chemical basis of morphogenesis) has been widely applied to a number of areas in developmental biology, where evolving cell and tissue structures are naturally observed. The related pattern formation models normally comprise a system of reaction-diffusion equations for interacting chemical species (morphogens), whose heterogeneous distribution in some spatial domain acts as a template for cells to form some kind of pattern or structure through, for example, differentiation or proliferation induced by the chemical pre-pattern. Here we develop a hybrid discrete-continuum modelling framework for the formation of cellular patterns via the Turing mechanism. In this framework, a stochastic individual-based model of cell movement and proliferation is combined with a reaction-diffusion system for the concentrations of some morphogens. As an illustrative example, we focus on a model in which the dynamics of the morphogens are governed by an activator-inhibitor system that gives rise to Turing pre-patterns. The cells then interact with the morphogens in their local area through either of two forms of chemically-dependent cell action: Chemotaxis and chemically-controlled proliferation. We begin by considering such a hybrid model posed on static spatial domains, and then turn to the case of growing domains. In both cases, we formally derive the corresponding deterministic continuum limit and show that that there is an excellent quantitative match between the spatial patterns produced by the stochastic individual-based model and its deterministic continuum counterpart, when sufficiently large numbers of cells are considered. This paper is intended to present a proof of concept for the ideas underlying the modelling framework, with the aim to then apply the related methods to the study of specific patterning and morphogenetic processes in the future. 
\end{abstract}

\section{Introduction}
\label{sect:intro}
\subsection{Turing's (pre-)pattern theory}
In 1952, Alan M.~Turing's seminal work `{\it The chemical basis of morphogenesis}' introduced the theory according to which the heterogeneous spatial distribution of some chemicals that regulate cellular differentiation, called ``morphogens'', acts as a template ({i.e.,} a pre-pattern) for cells to form different kinds of patterns or structures during the embryonic development of an organism~\cite{turingchemical}. In his work, Turing proposed a system of reaction-diffusion equations, with linear reaction terms, modelling the space-time dynamics of the concentrations of two morphogens as the basis for the formation of such pre-patterns. The system had stable homogenous steady states which were driven unstable by diffusion, resulting in spatially heterogeneous morphogen distributions, as long as the diffusion rate of one of the chemical was much larger (order 10) than the other. Twenty years after the publication of Turing's paper, in 1972 Alfred~Gierer and Hans~Meinhardt further developed this theory through the introduction of activator-inhibitor systems ({i.e.}, reaction-diffusion systems with nonlinear reaction terms) and the notion of ``short-range activation and long-range inhibition''~\cite{gierermeinhardt1972}. The application of this theory to many problems in developmental biology was discussed a further ten years later in 1982, in Meinhardt's book `{\em Models of Biological Pattern Formation}'~\cite{meinhardt1982}, shortly after the specific application of the theory to animal coat markings by James D.~Murray~\cite{murray1981}. \mbox{Turing (pre-)patterns }and resulting cellular patterns have now been discussed widely since their introduction and investigated through further mathematical modelling approaches.
\subsection{Mathematical exploration of cell pattern formation via the Turing mechanism} 
Several continuum models formulated as systems of partial differential equations (PDEs) have been used to study cell pattern formation via the Turing mechanism, in different spatial dimensions and on domains of various shapes and sizes. Generally, spatial domains can be static or growing to represent the growth of an organism over time. In~\cite{maini2019turing}, the authors highlighted that there can be a minimum domain size required for pattern formation to occur, and that when considering a growing domain Turing patterns generally become more complex. Multiple authors have analytically and numerically studied pattern formation on growing domains~\cite{arcuri1986pattern,chaplain2001spatio,crampin1999reaction,crampin2002pattern,kondo1995reaction,krause2019influence,krause2020one,maini2012turing,madzvamuse2015stability}. Specifically, in the case where chemotaxis of cells is included ({i.e.}, when cells move up the concentration gradient of the activator), various authors have considered pattern formation mediated by the Turing mechanism on exponentially growing domains~\cite{madzvamuse2003moving,painter1999stripe}. Along with spatial aspects of cellular patterning, temporal aspects can be considered, such as the role of time-delays in pattern formation. For instance, in~\cite{lee2011dynamics} the authors investigated Turing pattern formation on a morphogen-regulated growing domain where there was a delay in the signalling, gene expression and domain-growth processes.

Turing patterns can arise as stripes, spots (peaks of high density) or reverse-spots (troughs of low density) depending on the particular choice of parameter values and initial distributions of morphogens~\cite{murray2007mathematical}. The different scenarios leading to these three distinct classes of pre-patterns have been explored mathematically by various authors~\cite{ermentrout1991stripes,shoji2003stripes}. For example, in~\cite{meinhardt1989tailoring} the authors showed that, if there is a low level of production of the morphogens, striped patterns are formed by a wider range of parameter settings than spotted patterns. However, Turing patterns can be sensitive to small perturbations in the parameter values and the initial distributions of the morphogens, often leading to a discussion on the robustness of such patterns, or lack thereof~\cite{lee2011dynamics,maini2019turing}. In regard to a lack of robustness of the Turing mechanism to perturbations in the initial morphogen distributions, it has been found that incorporating stochastic aspects can improve robustness of pattern formation~\cite{maini2012turing}.

Discrete models and hybrid discrete-continuum models have also been used to describe cell pattern formation via the Turing mechanism in a range of biological and theoretical scenarios~\cite{chaturvedi2005multiscale,christley2007patterns,duggan2019generating,karig2018stochastic,kiskowski2004interplay,kondo2012turing,moreira2005pigment,okuda2018combining,volkening2015modelling}. In contrast to continuum models formulated as PDEs, such models permit the representation of biological processes at the level of single cells and account for possible stochastic variability in cell dynamics. This leads to greater adaptability and higher accuracy in the mathematical modelling of morphogenesis and pattern formation in biological systems~\cite{glen2019agent}. In particular, softwares like CompuCell~\cite{izaguirre2004compucell} and CompuCell3D~\cite{cickovski2005framework} have been employed to implement hybrid discrete-continuum models to investigate the interplay between gene regulatory networks and cellular processes ({e.g.}, haptotaxis, chemotaxis, cell adhesion and division) during morphogenesis. The three main components of models for cell pattern formation implemented using these softwares are: A Cellular Potts model for the dynamics of the cells and the extracellular matrix; a reaction-diffusion model for the dynamics of the diffusible morphogens; a combination of a state automaton and a set of ordinary differential equations to model the dynamics of gene regulatory networks. 
\subsection{A hybrid discrete-continuum approach to model cell pattern formation via the Turing mechanism} 
Here we develop a discrete-continuum modelling framework for the formation of cellular patterns via the Turing mechanism. In this framework, a reaction-diffusion system for the concentrations of some morphogens is combined with a stochastic individual-based (IB) model that tracks the dynamics of single cells. Such an IB model consists of a set of rules that describe cell movement and proliferation as a discrete-time branching random walk~\cite{hughes1995random}. 

A key advantage of this modelling framework is that it can be easily adapted to both static and growing spatial domains, thus covering a broad spectrum of applications. Moreover, the deterministic continuum limits of the IB models defined in this framework can be formally derived. Such continuum models are formulated as PDEs, which cannot capture phenomena that are driven by stochastic effects in the dynamics of single cells but are more amenable to analytical and numerical approaches. For instance, the numerical simulation of these models requires computational times that are typically much smaller than those required by the numerical exploration of the corresponding IB models for large cell numbers. Continuum models for spatial dynamics of living organisms have been derived from underlying discrete models through different mathematical methods in several previous works. Possible examples include the derivation of continuum models of chemotaxis from velocity-jump process~\cite{othmer2000diffusion,hillen2009user,othmer1988models,Painter2002,painter2003modelling} or from different types of random walks~\cite{alt1980biased,bubba2020discrete,burger2011continuous,stevens2000derivation,stevens1997aggregation}; the derivation of diffusion and nonlinear diffusion equations from random walks~\cite{champagnat2007invasion,chaplain2019bridging,inoue1991derivation,oelschlager1989derivation,othmer2002diffusion,penington2011building,penington2014interacting}, from systems of discrete equations of motion~\cite{baker2019free,byrne2009individual,lorenzi2019free,motsch2018short,murray2009discrete,murray2012classifying,oelschlager1990large}, from discrete lattice-based exclusion processes~\cite{binder2009exclusion,dyson2012macroscopic,fernando2010nonlinear,johnston2017co,johnston2012mean,landman2011myopic,lushnikov2008macroscopic,simpson2010cell} and from cellular automata~\cite{deroulers2009modeling,drasdo2005coarse,simpson2007simulating}; and the derivation of nonlocal models of cell-cell adhesion from position-jump processes~\cite{buttenschoen2018space}.

This paper is intended to be a proof of concept for the ideas underlying the modelling framework, with the aim to then apply the related methods to the study of specific patterning and morphogenetic processes, such as those discussed in~\cite{marcon2012turing,othmer2012experimental,othmer2009intersection} and references therein, in the future. 

\subsection{Contents of the paper} 
As an illustrative example, we focus on a hybrid discrete-continuum model in which the dynamics of the morphogens are governed by an activator-inhibitor system that gives rise to Turing pre-patterns. The cells then interact with the morphogens in their local area through either of two forms of chemically-dependent cell action: Chemotaxis and chemically-controlled proliferation. We begin by considering such a hybrid model posed on static spatial domains (see Section~\ref{sect:discrete}) and then turn to growing domains (see Section~\ref{sect:discretegrowing}). Using methods similar to those we have previously employed in~\cite{bubba2020discrete,chaplain2019bridging}, we formally derive the deterministic continuum limit of the model. When sufficiently large numbers of cells are considered, the results of numerical simulations demonstrate an excellent quantitative match between the spatial patterns produced by the stochastic IB model and its deterministic continuum counterpart.
Section~\ref{sect:conclusion} concludes the paper and provides a brief overview of possible research perspectives.

\section{Mathematical modelling of cell pattern formation on static domains}
\label{sect:discrete}
In this section, we illustrate our hybrid discrete-continuum modelling framework by developing a model for the formation of cellular patterns via the Turing mechanism on static spatial domains (see Section~\ref{sec:discmodstatdom}). The corresponding deterministic continuum model is provided (see Section~\ref{sec:contmodstatdom}) and results of numerical simulations of both models are presented (see Section~\ref{numsim1}). We report on numerical results demonstrating a good match between cellular patterns produced by the stochastic IB model and its deterministic continuum counterpart, in different spatial dimensions and biological scenarios, as well as on results showing the emergence of possible differences between the cell patterns produced by the two models for relatively low cell numbers.

\subsection{A hybrid discrete-continuum model}
\label{sec:discmodstatdom}
We let cells and morphogens be distributed across a $d$-dimensional static domain, with $d=1$ or $d=2$. In particular, we consider the case where the spatial domain is represented by the interval $[0,\ell]$ when $d=1$ or the square $[0,\ell] \times [0,\ell]$ when $d=2$, with $\ell \in \mathbb{R}^*_+$, where $\mathbb{R}^*_+$ is the set of positive real numbers not including zero. The position of the cells and the molecules of morphogens at time $t \in \mathbb{R}_+$ is modelled by the variable $x \in [0,\ell]$ when $d=1$ and by the vector ${\bf x}=(x,y) \in [0,\ell]^2$ when $d=2$.

We discretise the time variable $t$ as $t_{k} = k \tau$ with $k\in\mathbb{N}_{0}$ and the space variables $x$ and $y$ as $x_{i} = i \, \chi$ and $y_{j} = j \, \chi$ with $(i, j) \in [0,I]^2 \subset \mathbb{N}^2_0$, where $\tau \in \mathbb{R}^*_+$ and $\chi \in \mathbb{R}^*_+$ are the time- and space-step, respectively, and $I := 1 + \ceil[\Bigg]{\dfrac{\ell}{\chi}}$, where $\ceil{\cdot}$ denotes the ceiling function. Here, $\mathbb{N}_{0}$ is the set of natural numbers including zero. Throughout this section we use the notation ${\bf{i}} \equiv i$ and ${\bf x}_{\bf{i}} \equiv x_i$ when $d=1$, and ${\bf{i}} \equiv (i,j)$ and ${\bf x}_{\bf{i}} \equiv (x_i,y_j)$ when $d=2$. The concentrations of the morphogens at position ${\bf x}_{\bf i}$ and at time $t_k$ are modelled by the discrete, non-negative functions $u^{k}_{{\bf i}}$ and $v^{k}_{{\bf i}}$, and we denote by $n^{k}_{{\bf i}}$ the local cell density, which is defined as the number of cells at position ${\bf x}_{\bf{i}}$ and at time $t_k$, which is modelled by the dependent variable $N^{k}_{\bf i}\in\mathbb{N}_0$, divided by the size of the ${\bf i}^{th}$ site of the spatial grid, that is
\begin{equation}
\label{e:n}
n^k_{\bf i} := \dfrac{N^{k}_{\bf i}}{\chi^d}.
\end{equation}
We present here the model when $d=2$ but analogous considerations hold for $d=1$. 

\subsubsection{Dynamics of the morphogens}
The dynamics of $u^{k}_{{\bf i}}$ and $v^{k}_{{\bf i}}$ are governed by the following coupled system of difference equations 
\begin{equation}
\label{eq:discuv}
\begin{cases}
u^{k+1}_{{\bf i}} = u^{k}_{{\bf i}} + \dfrac{\tau D_u }{\chi^2} \left(\delta^2_{i} \ u^k_{{\bf i}} + \delta^2_{j} \ u^k_{{\bf i}}\right) + \tau \, P(u^{k}_{{\bf i}},v^{k}_{{\bf i}}),
\\\\
v^{k+1}_{{\bf i}} = v^{k}_{{\bf i}} + \dfrac{\tau D_v}{\chi^2} \left(\delta^2_{i} \ v^k_{{\bf i}} + \delta^2_{j} \ v^k_{{\bf i}}\right) + \tau \, Q(u^{k}_{{\bf i}},v^{k}_{{\bf i}}),
\end{cases}
\quad
(k,{\bf i}) \in \mathbb{N} \times (0,I)^2,
\end{equation}
subject to zero-flux boundary conditions. Here, $\delta^2_{i}$ is the second-order central difference operator on the lattice $\left\{x_i\right\}_i$ and $\delta^2_{j}$ is the second-order central difference operator on the lattice $\{y_j\}_j$, that is, 
\begin{equation}
\label{eq:socdo}
\delta^2_{i} u^k_{{\bf i}} := u^{k}_{(i+1,j)}+u^{k}_{(i-1,j)}-2\ u^{k}_{(i,j)} \quad \text{and} \quad \delta^2_{j} u^k_{{\bf i}} := u^{k}_{(i,j+1)}+u^{k}_{(i,j-1)}-2\ u^{k}_{(i,j)}.
\end{equation}
Moreover, $D_{u} \in \mathbb{R}^*_+$ and $D_{v} \in \mathbb{R}^*_+$ represent the diffusion coefficients of the morphogens and the functions $P(u^{k}_{{\bf i}},v^{k}_{{\bf i}})$ and $Q(u^{k}_{{\bf i}},v^{k}_{{\bf i}})$ are the rates of change of $u^{k}_{{\bf i}}$ and $v^{k}_{{\bf i}}$ due to local reactions.

The system of difference equations~\eqref{eq:discuv} is a standard discretisation of a generic reaction-diffusion system of the type that is commonly used to describe morphogen dynamics---see~\cite{maini2019turing,maini1997spatial} and references therein. The specific forms of the functions $P$ and $Q$ depend on the reaction kinetics involved in the biological problem at stake---we refer the interested reader to~\cite{maini2019turing,maini2012turing,murray1981} and references therein. We consider an activator-inhibitor system whereby $u^{k}_{{\bf i}}$ models the concentration of the activator while $v^{k}_{{\bf i}}$ models the concentration of the inhibitor. Hence, we let the functions $P$ and $Q$ satisfy the following standard assumptions for activator-inhibitor kinetics
\begin{equation}
\label{ass:defPQ}
\dfrac{\partial P}{\partial v} < 0 \quad \text{and} \quad \dfrac{\partial Q}{\partial u} > 0.
\end{equation}
In particular, in this paper we will focus on Schnakenberg kinetics~\cite{schnakenberg1979simple} and, therefore, the functions $P$ and $Q$ are given via the definitions~\eqref{schnackenbergA}. Moreover, we will assume
\begin{equation}
\label{ass:maxp}
0 < u^k_{\bf i} \leq u_{\rm max} \quad \text{and} \quad 0 < v^k_{\bf i} \leq v_{\rm max} \quad \forall \, (k, {\bf i}) \in \mathbb{N}_{0} \times [0,I]^2
\end{equation}
for some maximal concentrations $u_{\rm max} \in \mathbb{R}^*_+$ and $v_{\rm max} \in \mathbb{R}^*_+$.

\subsubsection{Dynamics of the cells}
\label{describeIB}
We consider a scenario where the cells proliferate ({i.e.}, divide and die) and can change their position according to a combination of undirected, random movement and chemotactic movement, which are seen as independent processes. This results in the following rules for the dynamic of the cells.

\paragraph{Mathematical modelling of undirected, random cell movement} We model undirected, random cell movement as a random walk with movement probability $\theta \in \mathbb{R}^*_+$, where $0<\theta\leq1$. In particular, for a cell on the lattice site ${\bf i}$, we define the probability of moving to one of the four neighbouring lattice sites in the von Neumann neighbourhood of ${\bf i}$ via undirected, random movement as $\dfrac{\theta}{4}$ ({i.e.,} there is an equal chance of moving to any of the four neighbouring sites). Therefore, the probability of not undergoing undirected, random movement is defined as $1 - \theta$ ({i.e.,} one minus the probability of movement). Furthermore, the spatial domain is assumed to be closed and, therefore, cell moves that require moving out of the spatial domain are not allowed. Under these assumptions, the probabilities of moving to the left and right sites via undirected, random movement are defined as
\begin{eqnarray}
\label{e:diffusionLR}
&\mathcal{T}^{k}_{{\rm L} (i,j)} := \dfrac{\theta}{4}, \quad \mathcal{T}^{k}_{{\rm R} (i,j)} := \dfrac{\theta}{4} \quad \text{for } (i,j)\in [1,I-1] \times [0,I], \nonumber \\\\
&\mathcal{T}^{k}_{{\rm L} (0,j)} := 0,\quad \mathcal{T}^{k}_{{\rm R} (I,j)} := 0 \quad \text{for } j \in [0,I], \nonumber
\end{eqnarray}
while the probabilities of moving to the lower and upper sites via undirected, random movement are defined as
\begin{eqnarray}
\label{e:diffusionDU}
&\mathcal{T}^{k}_{{\rm D} (i,j)} := \dfrac{\theta}{4}, \quad \mathcal{T}^{k}_{{\rm U} (i,j)} :=\dfrac{\theta}{4} \quad \text{for } (i,j)\in [0,I] \times [1,I-1], \nonumber \\\\
&\mathcal{T}^{k}_{{\rm D} (i,0)} := 0,\quad \mathcal{T}^{k}_{{\rm U} (i,I)} := 0 \quad \text{for } i \in [0,I]. \nonumber
\end{eqnarray}

\paragraph{Mathematical modelling of chemotactic cell movement} 
In line with~\cite{madzvamuse2003moving,painter1999stripe}, we let cells move up the concentration gradient of the activator via chemotaxis. Chemotactic cell movement is modelled as a biased random walk whereby the movement probabilities depend on the difference between the concentration of the activator at the site occupied by a cell and the concentration of the activator in the von Neumann neighbourhood of the cell's site. Moreover, as similarly done in the case of undirected, random cell movement, cell moves that require moving out of the spatial domain are not allowed. In particular, building on the modelling strategy presented in~\cite{bubba2020discrete}, for a cell on the lattice site ${\bf i}$ and at the time-step $k$, we define the probability of moving to the left and right sites via chemotaxis as
\begin{eqnarray}
\label{e:Jleftright}
& \mathcal{J}^{k}_{{\rm L} (i,j)} := \eta \, \dfrac{\left(u^{k}_{(i-1,j)}-u^{k}_{(i,j)}\right)_{+}}{4 u_{\rm max}}, \quad \mathcal{J}^{k}_{{\rm R} (i,j)} := \eta \, \dfrac{\left(u^{k}_{(i+1,j)}-u^{k}_{(i,j)}\right)_{+}}{4 u_{\rm max}} \quad \text{for } (i,j)\in [1,I-1] \times [0,I], \nonumber \\\\
& \mathcal{J}^{k}_{{\rm L} (0,j)} := 0, \quad \mathcal{J}^{k}_{{\rm R} (I,j)} := 0 \quad \text{for } j \in [0,I], \nonumber
\end{eqnarray}
while the probabilities of moving to the lower and upper sites via chemotaxis are defined as
\begin{eqnarray}
\label{e:Jdownup}
& \mathcal{J}^{k}_{{\rm D} (i,j)} := \eta \, \dfrac{\left(u^{k}_{(i,j-1)}-u^{k}_{(i,j)}\right)_{+}}{4 u_{\rm max}}, \quad \mathcal{J}^{k}_{{\rm U} (i,j)} := \eta \, \dfrac{\left(u^{k}_{(i,j+1)}-u^{k}_{(i,j)}\right)_{+}}{4 u_{\rm max}} \quad \text{for } (i,j)\in [0,I] \times [1,I-1], \nonumber \\\\
& \mathcal{J}^{k}_{{\rm D} (i,0)} := 0, \quad \mathcal{J}^{k}_{{\rm U} (i,I)} := 0 \quad \text{for } i \in [0,I]. \nonumber
\end{eqnarray}
Hence, the probability of not undergoing chemotactic movement is 
\begin{equation}
\label{e:stay}
1 - \left(\mathcal{J}^{k}_{{\rm L} {\bf i}} + \mathcal{J}^{k}_{{\rm R} {\bf i}} + \mathcal{J}^{k}_{{\rm D} {\bf i}} + \mathcal{J}^{k}_{{\rm U} {\bf i}}\right) \quad \text{for } {\bf i}\in [0,I]^2.
\end{equation}
Here, $(\cdot)_{+}$ denotes the positive part of $(\cdot)$ and the parameter $\eta \in \mathbb{R}_+$ with $0 \leq \eta \leq 1$, where $\mathbb{R}_+$ is the set of non-negative real numbers, is directly proportional to the chemotactic sensitivity of the cells. Notice that since relations~\eqref{ass:maxp} hold the quantities given via the definitions~\eqref{e:Jleftright}-\eqref{e:stay} are all between $0$ and $1$. 

\paragraph{Mathematical modelling of cell proliferation} We consider a scenario in which the cell population undergoes saturating growth. Hence, in line with~\cite{myerscough1998pattern}, we allow every cell to divide or die with probabilities that depend on a monotonically decreasing function of the local cell density. Moreover, building on the ideas presented in~\cite{sherratt1991mathematical}, we model chemically-controlled cell proliferation by letting the probabilities of cell division and death depend on the local concentrations of the activator and of the inhibitor. In particular, building upon the modelling strategy used in~\cite{chaplain2019bridging}, between time-steps $k$ and $k+1$, we let a cell on the lattice site ${\bf i}$ divide ({i.e.}, be replaced by two identical daughter cells that are placed on the lattice site ${\bf i}$) with probability
 \begin{equation} \label{pb}
 \mathcal{P}_b\left(n_{{\bf i}}^{k}, u_{{\bf i}}^{k}\right) := \tau \, \alpha_{n}\ \left(\psi(n_{{\bf i}}^{k})\right)_{+} \phi_u(u_{{\bf i}}^{k}),
 \end{equation}
die with probability 
 \begin{equation} \label{pd}
 \mathcal{P}_d\left(n_{{\bf i}}^{k}, u_{{\bf i}}^{k}\, v_{{\bf i}}^{k}\right) := \tau \, \left(\alpha_{n} \, \left(\psi(n_{{\bf i}}^{k})\right)_{-} \, \phi_u(u_{{\bf i}}^{k}) + \beta_n \, \phi_v(v_{{\bf i}}^{k})\right),
 \end{equation}
or remain quiescent ({i.e.}, do not divide nor die) with probability
 \begin{equation} \label{pq}
 \mathcal{P}_q\left(n_{{\bf i}}^{k}, u_{{\bf i}}^{k}, v_{{\bf i}}^{k} \right) := 1 - \tau \left(\alpha_{n}\ \left|\psi(n_{{\bf i}}^{k})\right| \, \phi_u(u_{{\bf i}}^{k}) + \beta_{n} \, \phi_v(v_{{\bf i}}^{k}) \right). 
 \end{equation}
Here, $(\cdot)_{+}$ and $(\cdot)_{-}$ denote the positive part and the negative part of $(\cdot)$. The parameters $\alpha_n \in \mathbb{R}^*_+$ and $\beta_n \in \mathbb{R}^*_+$ are, respectively, the intrinsic rates of cell division and cell death. Moreover, the function $\psi$ model the effects of saturating growth and, therefore, it is assumed to be such that
\begin{equation} \label{ass:psi}
\psi'(\cdot) < 0 \quad \mbox{and} \quad \psi(n_{\rm max}) = 0,
\end{equation}
 where $n_{\rm max} \in \mathbb{R}^*_+$ is the local carrying capacity of the cell population. Finally, the functions $\phi_u$ and $\phi_v$ satisfy the following assumptions 
\begin{equation} \label{ass:phi}
\phi_{u}(0)=1, \quad \phi'_{u}(\cdot) > 0 \quad \text{and} \quad \phi_{v}(0)=1, \quad \phi'_{v}(\cdot) > 0.
\end{equation}
Notice that we are implicitly assuming that the time-step $\tau$ is sufficiently small that $0 < \mathcal{P}_h < 1$ for all $h \in \{b,d,q \}$.

\subsection{Corresponding continuum model}
\label{sec:contmodstatdom}
Letting the time-step $\tau \to 0$ and the spdescribed in the definitions~\eqref{e:diffusionLR} and~\eqref{e:diffusionDU}ace-step $\chi \to 0$ in such a way that
\begin{equation}
\label{ass:quottozero}
\frac{\theta \, \chi^{2}}{2d \, \tau} \rightarrow D_{n} \quad \text{and}\quad \frac{\eta \, \chi^{2}}{2d \, \tau \, u_{\rm max}} \rightarrow C_{n} \quad \text{as } \tau \to 0, \; \chi \to 0,
\end{equation} 
using the formal method employed in~\cite{bubba2020discrete,chaplain2019bridging} it is possible to formally show (see Remark~\ref{remark1} in Appendix~\ref{app:derivation}) that the deterministic continuum counterpart of the stochastic IB model presented in Section~\ref{sec:discmodstatdom}, which is described via the system~\eqref{e:diffusionLR}~-~\eqref{ass:phi}, is given by the following PDE for the cell density $n(t,{\bf x})$
\begin{equation}
\label{eq:PDEn}
\partial_t n - \nabla_{{\bf x}} \cdot \left(D_{n} \, \nabla_{{\bf x}} n - C_{n} \, n \, \nabla_{{\bf x}} u \right) = \Big(\alpha_{n}\ \psi(n) \ \phi_u(u) - \beta_n \ \phi_v(v) \Big)\ n, \quad (t,{\bf x}) \in \mathbb{R}^*_+ \times (0,\ell)^d
\end{equation}
subject to zero-flux boundary conditions. Here, $D_{n} \in \mathbb{R}^*_+$ defined via conditions~\eqref{ass:quottozero} is the diffusion coefficient ({i.e.}, the motility) of the cells, while $C_{n} \in \mathbb{R}_+$ defined via conditions~\eqref{ass:quottozero} represents the chemotactic sensitivity of the cells to the activator. In equation~\eqref{eq:PDEn}, the concentration of the activator $u(t,{\bf x})$ and the concentration of the inhibitor $v(t,{\bf x})$ are governed by the continuum counterpart of the system of difference equations~\eqref{eq:discuv} subject to zero-flux boundary conditions, that is, the following system of PDEs complemented with zero-flux boundary conditions
\begin{equation}
\label{eq:PDEuv}
\begin{cases}
\displaystyle{\partial_t u - D_{u} \, \Delta_{{\bf x}} u = P(u,v),}\\
\displaystyle{\partial_t v - D_{v} \, \Delta_{{\bf x}} u = Q(u,v)},
\end{cases}
\quad (t,{\bf x}) \in \mathbb{R}^*_+ \times (0,\ell)^d,
\end{equation}
which can be formally obtained by letting $\tau \to 0$ and $\chi \to 0$ in system~\eqref{eq:discuv}.

\subsection{Numerical simulations}
\label{numsim1}
In this section, we carry out a systematic quantitative comparison between the results of numerical simulations of the hybrid model presented in Section~\ref{sec:discmodstatdom} and numerical solutions of the corresponding continuum model given in Section~\ref{sec:contmodstatdom}, both in one and in two spatial dimensions. All simulations are performed in {\sc{Matlab}} and the final time of simulations is chosen such that the concentrations of morphogens and the cell density are at numerical equilibrium at the end of simulations.

\subsubsection{Summary of the set-up of numerical simulations}
\paragraph{Dynamics of the morphogens} We consider the case where the functions $P$ and $Q$ that model the rates of change of the concentrations of the morphogens are defined according to Schnakenberg kinetics~\cite{schnakenberg1979simple}, that is,
\begin{equation} 
\label{schnackenbergA}
P(u,v) := \alpha_{u} - \beta \, u + \gamma \, u^{2} \, v, \quad Q(u,v) := \alpha_{v} - \gamma \, u^{2} \, v
\end{equation}
where $\alpha_{u},\, \alpha_{v},\, \beta,\, \gamma \in \mathbb{R}^*_+$. The system of difference equations~\eqref{eq:discuv} and the system of PDEs~\eqref{eq:PDEuv} complemented with definitions~\eqref{schnackenbergA} and subject to zero-flux boundary conditions are known to exhibit Turing pre-patterns. The conditions required for such patterns to emerge have been extensively studied in previous works and, therefore, are omitted here---the interested reader is referred to~\cite{maini2019turing} and references therein. We choose parameter values such that Turing pre-patterns arise from the perturbation of homogeneous initial distributions of the morphogens. A complete description of the set-up of numerical simulations is given in Appendix~\ref{app:sta}. 

\paragraph{Dynamics of the cells} We focus on the case where the cell population undergoes logistic growth and, therefore, we define the function $\psi$ in equations~\eqref{pb}--\eqref{pq} and equation~\eqref{eq:PDEn} as
 \begin{equation} \label{def:psi}
\psi(n) := \left(1 - \dfrac{n}{n_{\rm max}}\right).
\end{equation}
Moreover, we consider two scenarios corresponding to two alternative forms of chemically-dependent cell action. In the first scenario, there is no chemotaxis---{i.e.}, we assume $\eta=0$ in definitions~\eqref{e:Jleftright} and~\eqref{e:Jdownup}, which implies that $C_{n}=0$ in equation~\eqref{eq:PDEn}---and the cells interact with the morphogens in their local area through chemically-controlled proliferation. In particular, we use the following definitions of the functions $\phi_u$ and $\phi_v$ in equations~\eqref{pb}--\eqref{pq} and equation~\eqref{eq:PDEn}
\begin{equation} \label{def:phi}
\phi_u(u) := 1 + \dfrac{u}{u_{\rm max}} \quad \text{and} \quad \phi_v(v) := 1 + \dfrac{v}{v_{\rm max}}.
\end{equation}
\indent In the second scenario, chemotaxis up the concentration gradient of the activator occurs---{i.e.}, we assume $\eta>0$, which implies that $C_{n}>0$---but cell division and death are not regulated by the morphogens---{i.e.}, we assume
\begin{equation} \label{def:phie1}
\phi_u(u) \equiv 1 \quad \text{and} \quad \phi_v(v) \equiv 1.
\end{equation} 

In both scenarios, we let the initial cell distribution be homogeneous and, given the values of the parameters chosen to carry out numerical simulations of the IB model, we use the following definitions
\begin{equation}
\label{ass:quottozeronum}
D_{n} := \frac{\theta \, \chi^{2}}{2 \, d \, \tau} \quad \text{and}\quad C_n := \frac{\eta \, \chi^{2}}{2 \, d \, \tau \, u_{\rm max}}
\end{equation} 
so that conditions~\eqref{ass:quottozero} are met. A complete description of the set-up of numerical simulations is given in Appendix~\ref{app:sta}.

\subsubsection{Main results of numerical simulations}
\label{sec:results_stat}

\paragraph{Dynamics of the morphogens} The plots in the top rows of Figures~\ref{fig1Dproliferation} and~\ref{fig1Dchemotaxis} and in the Supplementary Figure~\ref{fig2Dprepattern} summarise the dynamics of the continuum concentrations of morphogens $u(t,{\bf x})$ and $v(t,{\bf x})$ obtained by solving numerically the system of PDEs~\eqref{eq:PDEuv} subject to zero-flux boundary conditions. Identical results hold for the discrete morphogen concentrations $u_{\bf i}^k$ and $v_{\bf i}^k$ obtained by solving the system  of difference equations~\eqref{eq:discuv} (results not shown). These plots demonstrate that in the case where the reaction terms $P$ and $Q$ are described via the definitions~\eqref{schnackenbergA}, under the parameter setting considered here, Turing pre-patterns arise from perturbation of homogeneous initial distributions of the morphogens. Such pre-patterns consist of spots of activator and inhibitor, whereby maximum points of the concentration of activator coincide with minimum points of the concentration of inhibitor, and vice versa. 
\vspace{-0.5cm}
\paragraph{Dynamics of the cells in the presence of chemically-controlled cell proliferation} The plots in the bottom row of Figure~\ref{fig1Dproliferation} and the plots in Figure~\ref{fig2Dproliferation} summarise the dynamics of the cell density in the case where there is no chemotaxis and chemically-controlled cell proliferation occurs---{i.e.}, when $\eta=0$, $C_n=0$, and the functions $\phi_u$ and $\phi_v$ are given via the definitions~\eqref{def:phi}. Since a larger concentration of the activator corresponds to a higher cell division rate and a smaller concentration of the inhibitor corresponds to a lower cell death rate, we observe the formation of cellular patterns consisting of spots of cells centred at the same points as the spots of activator. These plots demonstrate that there is an excellent quantitative match between the discrete cell density $n^k_{\bf i}$ given by equation~\eqref{e:n}, with $N^k_{\bf i}$ obtained through computational simulations of the IB model, and the continuum cell density $n(t,{\bf x})$ obtained by solving numerically the PDE~\eqref{eq:PDEn} subject to zero-flux boundary conditions, both in one and in two spatial dimensions. 

\vspace{-0.5cm} 
\paragraph{Dynamics of the cells in the presence of chemotaxis} The plots in the bottom row of Figure~\ref{fig1Dchemotaxis} and the plots in Figure~\ref{fig2Dchemotaxis} summarise the dynamics of the cell density in the case where cell proliferation is not regulated by the morphogens and chemotactic movement of the cells up the concentration gradient of the activator occurs---{i.e.}, when the functions $\phi_u$ and $\phi_v$ are given via the definitions~\eqref{def:phie1}, $\eta>0$ and $C_n>0$. Since the cells sense the concentration of the activator and move up its gradient, cellular patterns consisting of spots of cells centred at the same points as the spots of the activator are formed. Compared to the case of chemically-controlled cell proliferation, in this case the spots of cells are smaller and characterised by a larger cell density ({i.e.}, cells are more densely packed). There is again an excellent quantitative match between the discrete cell density $n^k_{\bf i}$ given by equation~\eqref{e:n}, with $N^k_{\bf i}$ obtained through computational simulations of the IB model, and the continuum cell density $n(t,{\bf x})$ obtained by solving numerically the PDE~\eqref{eq:PDEn} subject to zero-flux boundary conditions, both in one and in two spatial dimensions.

 \begin{figure}[h!]
 	\centering
\includegraphics[width=\textwidth]{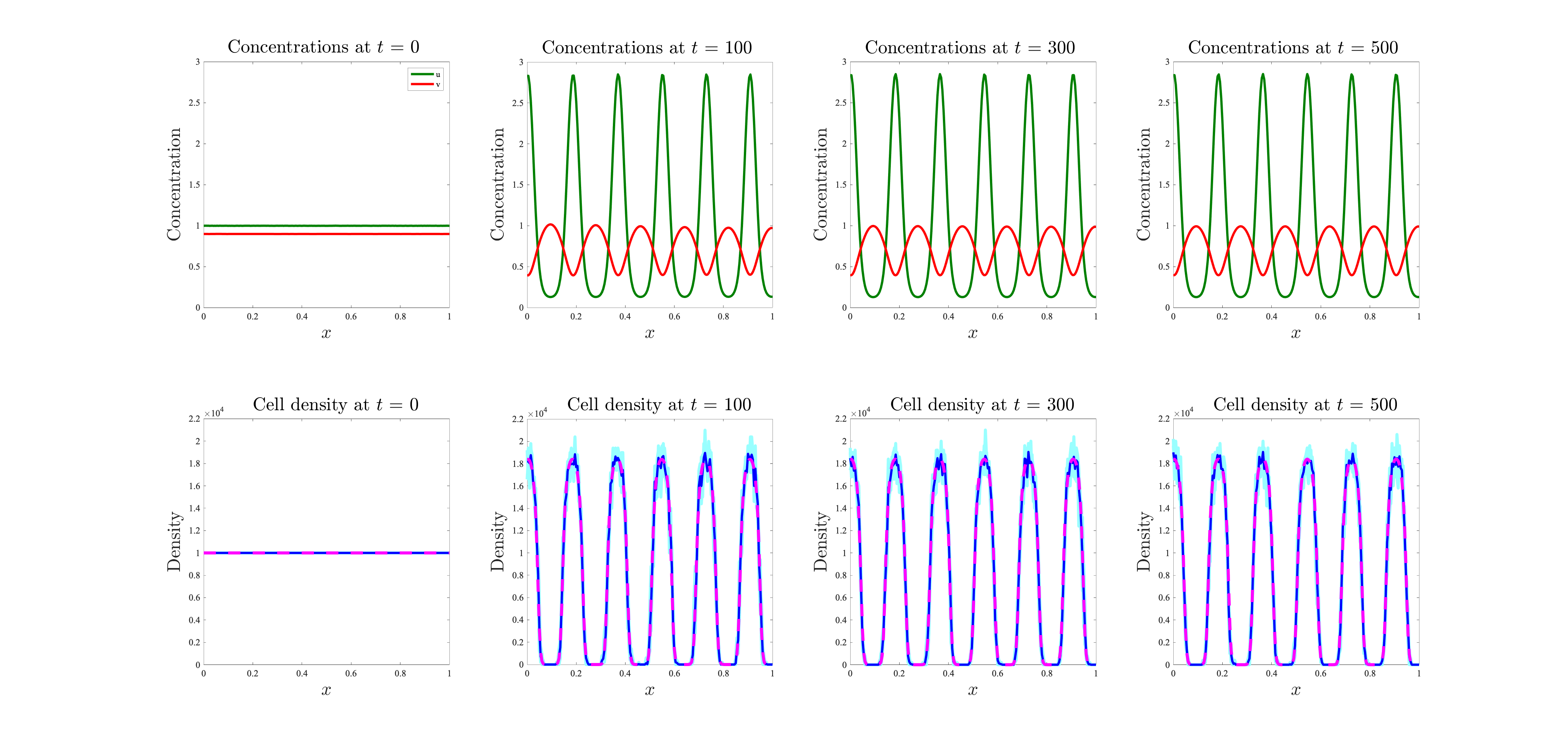}
\caption{{Results of numerical simulations on a one-dimensional static domain in the presence of chemically-controlled cell proliferation}. {(Top row)} Plots of the concentrations of morphogens at four consecutive time instants. The green lines highlight the concentration of activator $u(t,x)$ and the red lines highlight the concentration of inhibitor $v(t,x)$ obtained by solving numerically the system of PDEs~\eqref{eq:PDEuv} for $d=1$ complemented with the definitions~\eqref{schnackenbergA} and subject to zero-flux boundary conditions. (Bottom row) Comparison between the discrete cell density $n^k_i$ obtained by averaging the results of computational simulations of the IB model (solid dark blue lines) and the continuum cell density $n(t,x)$ obtained by solving numerically the PDE~\eqref{eq:PDEn} for $d=1$ subject to zero-flux boundary conditions (pink dashed lines), at four consecutive time instants. Here, $\eta=0$, $C_n=0$, and the functions $\phi_u$ and $\phi_v$ are given by definitions~\eqref{def:phi}. We additionally set the initial cell density $n_{i}^{0}= 10^4$ for all $i$. The results from the IB model correspond to the average over five realisations of the underlying branching random walk, with the results from each realisation plotted in pale blue to demonstrate the robustness of the results obtained. A complete description of the set-up of numerical simulations is given in Appendix~\ref{app:sta}.\label{fig1Dproliferation}}
\end{figure}

\begin{figure}[h!]
\centering
\includegraphics[width=\textwidth]{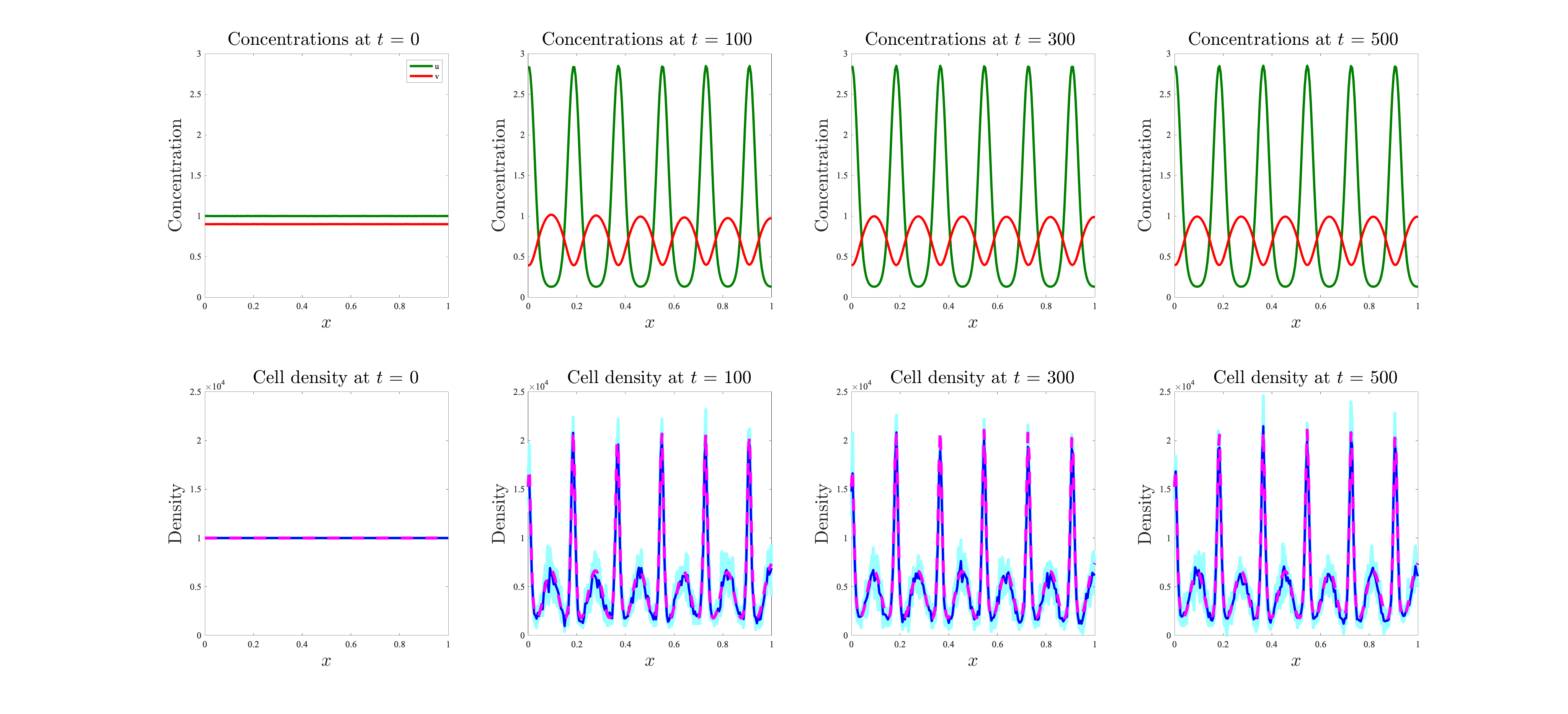}
\caption{{ Results of numerical simulations on a one-dimensional static domain in the presence of chemotaxis}. (Top row) Plots of the concentrations of morphogens at four consecutive time instants. The green lines highlight the concentration of activator $u(t,x)$ and the red lines highlight the concentration of inhibitor $v(t,x)$ obtained by solving numerically the system of PDEs~\eqref{eq:PDEuv} for $d=1$, complemented with the definitions~\eqref{schnackenbergA} and subject to zero-flux boundary conditions. (Bottom row) Comparison between the discrete cell density $n^k_i$ obtained by averaging the results of computational simulations of the IB model (solid dark blue lines) and the continuum cell density $n(t,x)$ obtained by solving numerically the PDE~\eqref{eq:PDEn} for $d=1$ subject to zero-flux boundary conditions (pink dashed lines), at four consecutive time instants. Here, $\eta>0$, $C_n>0$, and the functions $\phi_u$ and $\phi_v$ are described through the definitions~\eqref{def:phie1}. We additionally set the initial cell density $n_{i}^{0}= 10^4$ for all $i$. The results from the IB model correspond to the average over five realisations of the underlying branching random walk, with the results from each realisation plotted in pale blue to demonstrate the robustness of the results obtained. A complete description of the set-up of numerical simulations is given in Appendix~\ref{app:sta}.\label{fig1Dchemotaxis}}
\end{figure}

 \begin{figure}[h!]
 	\centering
\includegraphics[width=\textwidth]{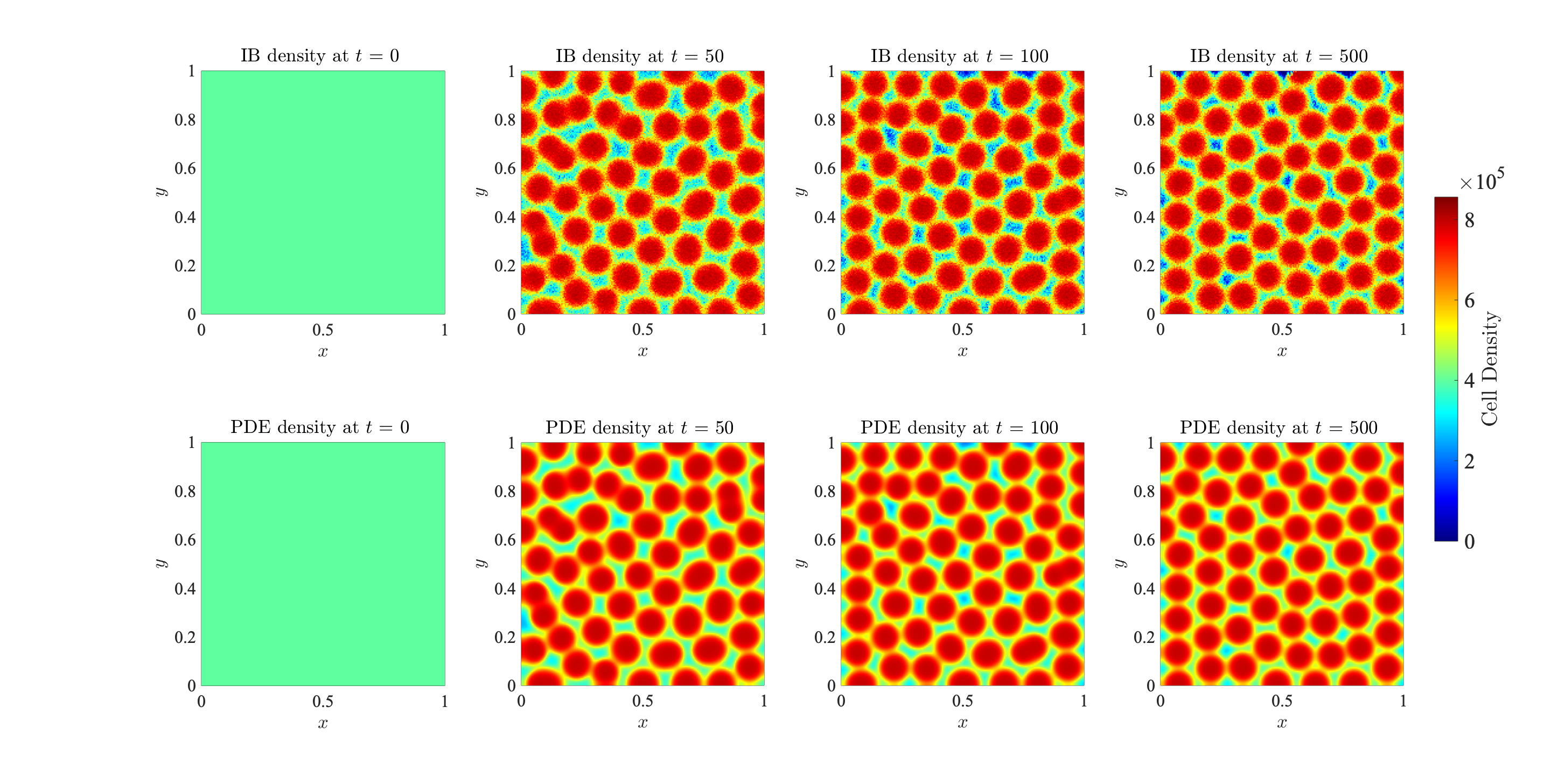}
\caption{{Results of numerical simulations on a two-dimensional static domain in the presence of chemically-controlled cell proliferation}. Comparison between the discrete cell density $n^k_{\bf i}$ obtained by averaging the results of computational simulations of the IB model (top row) and the continuum cell density $n(t,{\bf x})$ obtained by solving numerically the PDE~\eqref{eq:PDEn} for $d~=~2$ subject to zero-flux boundary conditions (bottom row), at four consecutive time instants. Here, $\eta=0$, $C_n=0$, and the functions $\phi_u$ and $\phi_v$ are given by definitions~\eqref{def:phi}. We additionally set the initial cell density $n_{\bf i}^{0}=4\times10^5$ for all ${\bf i}$. The results from the IB model correspond to the average over five realisations of the underlying branching random walk. The plots of the corresponding morphogen concentrations are displayed in the Supplementary Figure~\ref{fig2Dprepattern}. A complete description of the set-up of numerical simulations is given in Appendix~\ref{app:sta}.\label{fig2Dproliferation}}
\end{figure}

 \begin{figure}[h!]
 	\centering
\includegraphics[width=\textwidth]{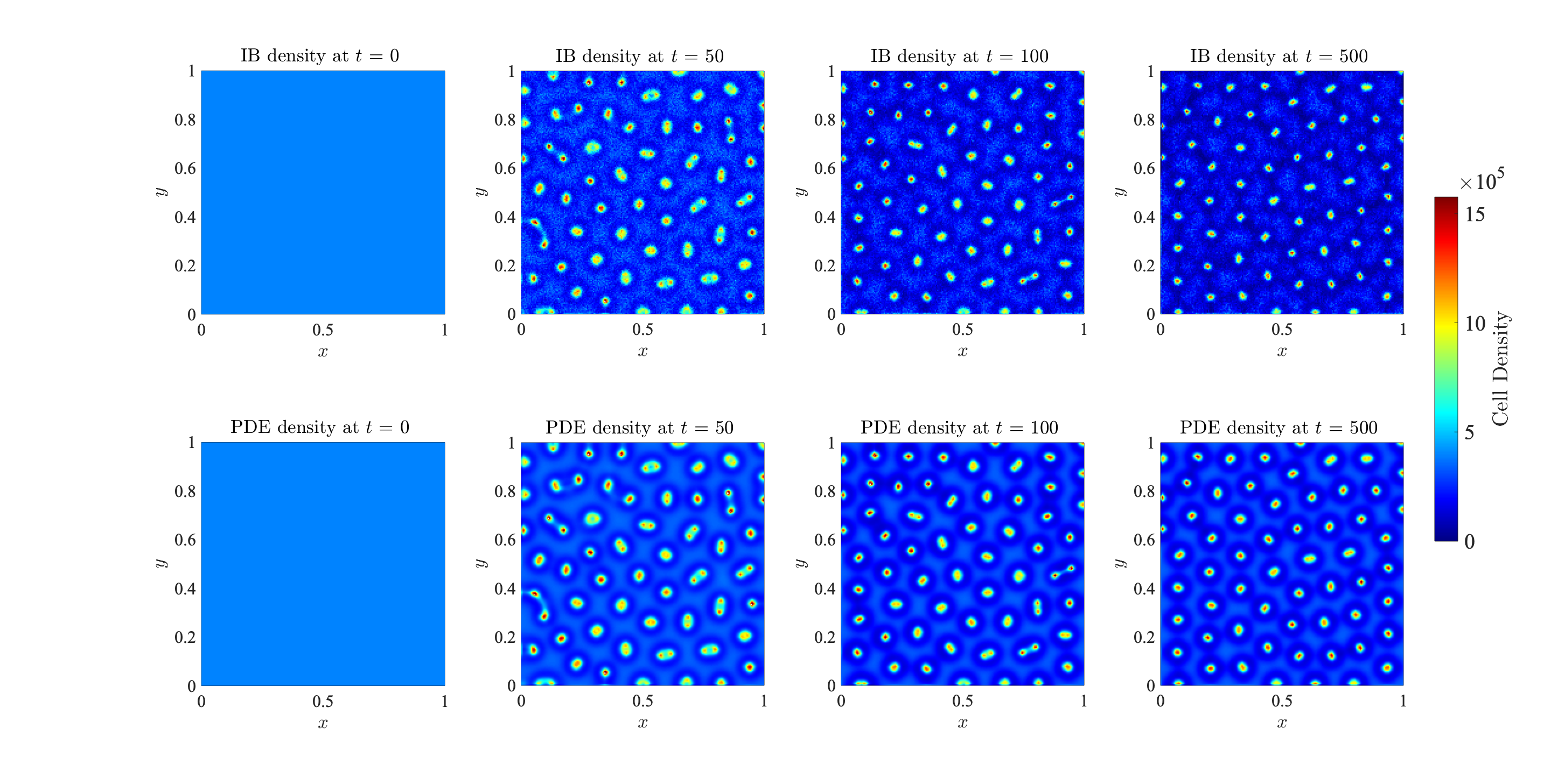}
\caption{{Results of numerical simulations on a two-dimensional static domain in the presence of chemically-controlled cell proliferation}. Comparison between the discrete cell density $n^k_{\bf i}$ obtained by averaging the results of computational simulations of the IB model (top row) and the continuum cell density $n(t,{\bf x})$ obtained by solving numerically the PDE~\eqref{eq:PDEn} for $d~=~2$ subject to zero-flux boundary conditions (bottom row), at four consecutive time instants. Here, $\eta>0$, $C_n>0$, and the functions $\phi_u$ and $\phi_v$ are described through the definitions~\eqref{def:phie1}. We additionally set the initial cell density $n_{\bf i}^{0}=4\times10^5$ for all ${\bf i}$. The results from the IB model correspond to the average over five realisations of the underlying branching random walk. The plots of the corresponding morphogen concentrations are displayed in the Supplementary Figure~\ref{fig2Dprepattern}. A complete description of the set-up of numerical simulations is given in Appendix~\ref{app:sta}.\label{fig2Dchemotaxis}}
\end{figure}
\begin{figure}[h!]
	\centering
\includegraphics[width=\textwidth]{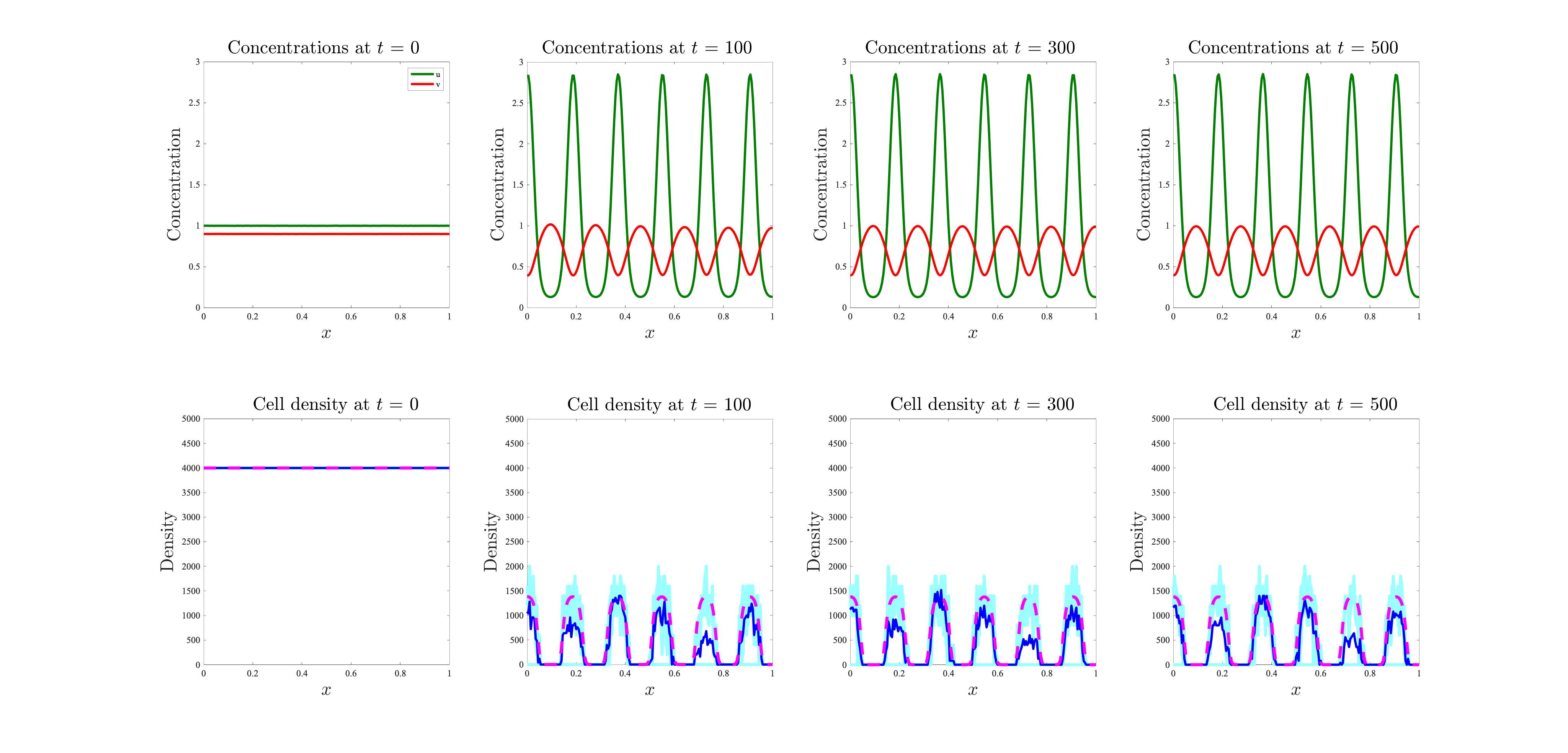}
\caption{{Emergence of possible differences between cell patterns produced by the IB model and the continuum model for low cell numbers on a one-dimensional static domain}. (Top row) Plots of the concentrations of morphogens at four consecutive time instants. The green lines highlight the concentration of activator $u(t,x)$ and the red lines highlight the concentration of inhibitor $v(t,x)$ obtained by solving numerically the system of PDEs~\eqref{eq:PDEuv} for $d=1$, complemented with the definitions~\eqref{schnackenbergA} and subject to zero-flux boundary conditions. (Bottom row) Comparison between the discrete cell density $n^k_i$ obtained by averaging the results of computational simulations of the IB model (solid dark blue lines) and the continuum cell density $n(t,x)$ obtained by solving numerically the PDE~\eqref{eq:PDEn} for $d=1$ subject to zero-flux boundary conditions (pink dashed lines), at four consecutive time instants. Here, $\eta=0$, $C_n=0$, and the functions $\phi_u$ and $\phi_v$ are given by definitions~\eqref{def:phi}. We additionally set the initial cell density $n_{i}^{0}=4\times10^3$ for all $i$. The results from the IB model correspond to the average over five realisations of the underlying branching random walk, with the results from each realisation plotted in pale blue. The parameter setting is the same as that of Figure~\ref{fig1Dproliferation} but with a smaller initial cell density and a smaller local carrying capacity of the cell population $n_{\rm max}$. A complete description of the set-up of numerical simulations is given in Appendix~\ref{app:sta}.\label{fig1Dproliferation_lowcell}}
\end{figure}
\begin{figure}[h!]
 	\centering
\includegraphics[width=\textwidth]{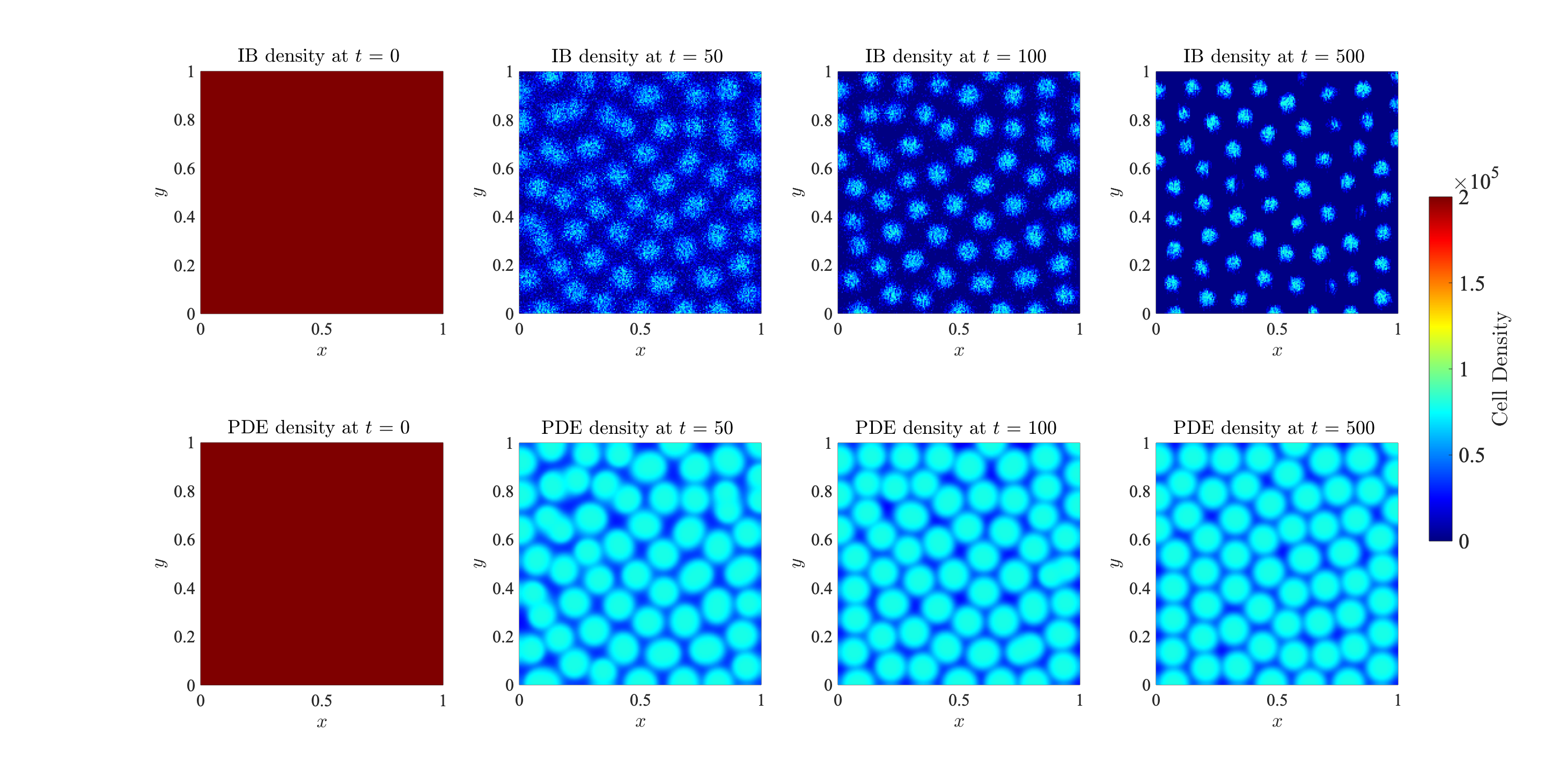}
\caption{{Emergence of possible differences between cell patterns produced by the IB model and the continuum model for low cell numbers on a two-dimensional static domain}. Comparison between the discrete cell density $n^k_{\bf i}$ obtained by averaging the results of computational simulations of the IB model (top row) and the continuum cell density $n(t,{\bf x})$ obtained by solving numerically the PDE~\eqref{eq:PDEn} for $d=2$ subject to zero-flux boundary conditions (bottom row), at four consecutive time instants. Here, $\eta=0$, $C_n=0$, and the functions $\phi_u$ and $\phi_v$ are given by definitions~\eqref{def:phi}. We additionally set the initial cell density $n_{\bf i}^{0}=4\times10^5$ for all ${\bf i}$. The results from the IB model correspond to the average over five realisations of the underlying branching random walk. The plots of the corresponding morphogen concentrations are displayed in the Supplementary Figure~\ref{fig2Dprepattern}. The parameter setting is the same as that of Figure~\ref{fig2Dproliferation} but with a smaller initial cell density and a smaller local carrying capacity of the cell population $n_{\rm max}$. A complete description of the set-up of numerical simulations is given in Appendix~\ref{app:sta}.\label{fig2Dproliferation_lowcell}}

\end{figure}

\vspace{-0.5cm}
\paragraph{Emergence of possible differences between cell patterns produced by the IB model and the continuum model} In all cases discussed so far we have observed excellent agreement between the dynamics of the discrete cell density obtained through computational simulations of the stochastic IB model and the continuum cell density obtained by solving numerically the corresponding deterministic continuum model. However, we expect possible differences between the two models to emerge in the presence of low cell numbers. In order to investigate this, we carried out numerical simulations of the IB model and the PDE model for the case where cells undergo chemically-controlled cell proliferation, considering either lower initial cell densities along with lower values of the local carrying capacity of the cell population $n_{\rm max}$ or higher rates of cell death $\beta_n$, which correspond to lower saturation values of the local cell density. The plots in the bottom row of Figure~\ref{fig1Dproliferation_lowcell} and in Figure~\ref{fig2Dproliferation_lowcell} summarise the dynamics of the cell density for relatively small initial cell numbers and local carrying capacities. These plots show that differences between the discrete cell density $n^k_{\bf i}$ given by equation~\eqref{e:n}, with $N^k_{\bf i}$ obtained through computational simulations of the IB model, and the continuum cell density $n(t,{\bf x})$, obtained by solving numerically the PDE~\eqref{eq:PDEn} subject to zero-flux boundary conditions, can emerge both in one and in two spatial dimensions. Analogous considerations hold for the case in which higher rates of cell death $\beta_n$ are considered (results not shown).

\section{Mathematical modelling of cell pattern formation on growing domains}
\label{sect:discretegrowing}
So far, we have considered the case of static spatial domains. However, in many biological problems the formation of cellular patterns occurs on growing spatial domains, for example, in morphogenesis and embryogenesis \cite{kondo1995reaction,krause2019influence,krause2020one,maini2012turing}. Therefore, in this section, we extend the hybrid model developed in the previous section to the case of growing spatial domains (see Section~\ref{hybrid_growing}). We consider both the case of uniform domain growth and the case of apical growth ({i.e.}, the case where domain growth is restricted to a region located toward the boundary). Similarly to the previous section, the deterministic continuum limit of the model is provided (see Section~\ref{continuum_growing}) and the results of numerical simulations demonstrating a good match between the cellular patterns produced by the stochastic IB model and its deterministic continuum counterpart are presented (see Section~\ref{numsim2}).

\subsection{A hybrid-discrete continuum model}
\label{hybrid_growing}
Building upon the modelling framework presented in the previous section, we let cells and morphogens be distributed across a $d$-dimensional growing domain represented by the interval $[0,\mathcal{L}(t)]$ when $d=1$ and the square $[0,\mathcal{L}(t)]^2$ when $d=2$. The real, positive function $\mathcal{L}(t)$, with $\mathcal{L}(\cdot) \geq 1$, determines the growth of the right-hand and upper boundary of the spatial domain ({i.e.}, we consider the case where the domain grows equally in both the $x$ and $y$ directions). In analogy with the previous section, we use the notation $x \in [0,\mathcal{L}(t)]$ and ${\bf x} = (x,y) \in [0,\mathcal{L}(t)]^2$. Moreover, we make the change of variables
\begin{equation}
\label{changeofv}
x \mapsto \hat{x} \quad \text{and}\quad {\bf x} \mapsto \hat{{\bf x}} = (\hat{x},\hat{y}) \quad \text{with} \quad \hat{x} := \frac{x}{\mathcal{L}(t)}, \;\; \hat{y} := \frac{y}{\mathcal{L}(t)}
\end{equation}
which allows one to describe the spatial position of the cells and the molecules of morphogens by means of the variable $\hat{x}\in[0,1]$ when $d=1$ and the vector $\hat{{\bf x}} = (\hat{x},\hat{y}) \in [0,1]^2$ when $d=2$~\cite{crampin1999reaction,lolasthesis}. 
We then discretise the time variable $t$ and space variables $\hat x$ and $\hat y$ in a similar way to the static domain case, as described in Section~\ref{sec:discmodstatdom}. Throughout this section we use the notation ${\bf{i}} \equiv i$ and $\hat {\bf x}_{\bf{i}} \equiv \hat x_i$ when $d=1$, and ${\bf{i}} \equiv (i,j)$ and $\hat {\bf x}_{\bf{i}} \equiv (\hat x_i, \hat y_j)$ when $d=2$. We also use the notation $\mathcal{L}_k = \mathcal{L}(t_k)$. The concentrations of the morphogens at position $\hat {\bf x}_{\bf i}$ and at time $t_k$ are modelled by the discrete, non-negative functions $u^{k}_{{\bf i}}$ and $v^{k}_{{\bf i}}$, and we denote by $n^{k}_{{\bf i}}$ the local cell density, which is defined via equation~\eqref{e:n}. As in the case of static domains, we present the model for $d=2$ but analogous considerations hold for $d=1$. 

\subsubsection{Dynamics of the morphogens}
The dynamics of $u^{k}_{{\bf i}}$ and $v^{k}_{{\bf i}}$ are governed by the following coupled system of difference equations 
\begin{equation}
\label{eq:discuvgro}
\begin{cases}
u^{k+1}_{{\bf i}} = u^{k}_{{\bf i}} + \dfrac{\tau D_u}{\mathcal{L}^2_k \, \chi^2} \left(\delta^2_{i} \ u^k_{{\bf i}} + \delta^2_{j} \ u^k_{{\bf i}}\right) + \tau \, P(u^{k}_{{\bf i}},v^{k}_{{\bf i}}) - g_{{\bf i}}(u^{k}_{{\bf i}}, \mathcal{L}_{k}),
\\\\
v^{k+1}_{{\bf i}} = v^{k}_{{\bf i}} + \dfrac{\tau D_v}{\mathcal{L}^2_k \, \chi^2} \left(\delta^2_{i} \ v^k_{{\bf i}} + \delta^2_{j} \ v^k_{{\bf i}}\right) + \tau \, Q(u^{k}_{{\bf i}},v^{k}_{{\bf i}}) - g_{{\bf i}}(v^{k}_{{\bf i}}, \mathcal{L}_{k}),
\end{cases}
\quad 
(k,{\bf i}) \in \mathbb{N} \times (0,I)^2,
\end{equation}
subject to zero-flux boundary conditions. Here, $\delta^2_{i}$ is the second-order central difference operator on the lattice $\left\{\hat x_{i}\right\}_{i}$ and $\delta^2_{j}$ is the second-order central difference operator on the lattice $\{\hat y_{j}\}_{j}$, which are provided in definitions~\eqref{eq:socdo}. Moreover, $D_{u} \in \mathbb{R}^*_+$ and $D_{v} \in \mathbb{R}^*_+$ represent the diffusion coefficients of the morphogens, which are rescaled by $\mathcal{L}^2_k$ for consistency with the change of variables~\eqref{changeofv}, and the functions $P(u^{k}_{{\bf i}},v^{k}_{{\bf i}})$ and $Q(u^{k}_{{\bf i}},v^{k}_{{\bf i}})$ are the rates of change of $u^{k}_{{\bf i}}$ and $v^{k}_{{\bf i}}$ due to local reactions, which satisfy conditions~\eqref{ass:defPQ}~and~\eqref{ass:maxp}, as in the case of static domains. Finally, the last terms on the right-hand sides of system~\eqref{eq:discuvgro} represent the rates of change of the concentrations of morphogens due to variation in the size of the domain. In the case of uniform domain growth, the following definition holds~\cite{crampin1999reaction,crampin2002pattern} 
\begin{equation}
\label{def:unifdg}
g_{{\bf i}}(w^{k}_{{\bf i}}, \mathcal{L}_{k}) \equiv g(w^{k}_{{\bf i}}, \mathcal{L}_{k}) := d \, w^{k}_{{\bf i}} \, \dfrac{\mathcal{L}_{k+1} - \mathcal{L}_k}{\mathcal{L}_k}.
\end{equation}
Equation~\eqref{def:unifdg} captures the effects of dilution of the concentrations of the morphogens due to local volume changes of the spatial domain~\cite{crampin1999reaction,krause2019influence}. On the other hand, when apical growth of the domain occurs one has~\cite{crampin2002pattern,lolasthesis}
\begin{equation}
\label{def:apicaldg}
g_{{\bf i}}(w^{k}_{{\bf i}}, \mathcal{L}_{k}) := \left[i \, \left(w^{k}_{i+1,j} - w^{k}_{i,j}\right) + j \, \left(w^{k}_{i,j+1} - w^{k}_{i,j}\right)\right] \, \dfrac{\mathcal{L}_{k+1} - \mathcal{L}_k}{\mathcal{L}_k}.
\end{equation}

\subsubsection{Dynamics of the cells}
\label{describeIBgro}
Under the change of variables~\eqref{changeofv}, the dynamics of the cells in the IB model is governed by rules analogous to those described in Section~\ref{describeIB} for the case of static domains. In summary, definitions~\eqref{e:diffusionLR} and~\eqref{e:diffusionDU} are modified as
\begin{eqnarray}
\label{e:diffusionLRgro}
&\mathcal{T}^{k}_{{\rm L} (i,j)} := \dfrac{\theta}{4 \mathcal{L}^2_k}, \quad \mathcal{T}^{k}_{{\rm R} (i,j)} := \dfrac{\theta}{4 \mathcal{L}^2_k} \quad \text{for } (i,j)\in [1,I-1] \times [0,I], \nonumber \\\\
&\mathcal{T}^{k}_{{\rm L} (0,j)} := 0,\quad \mathcal{T}^{k}_{{\rm R} (I,j)} := 0 \quad \text{for } j \in [0,I], \nonumber
\end{eqnarray}
\begin{eqnarray}
\label{e:diffusionDUgro}
&\mathcal{T}^{k}_{{\rm D} (i,j)} := \dfrac{\theta}{4 \mathcal{L}^2_k}, \quad \mathcal{T}^{k}_{{\rm U} (i,j)} :=\dfrac{\theta}{4 \mathcal{L}^2_k} \quad \text{for } (i,j)\in [0,I] \times [1,I-1], \nonumber \\\\
&\mathcal{T}^{k}_{{\rm D} (i,0)} := 0,\quad \mathcal{T}^{k}_{{\rm U} (i,I)} := 0 \quad \text{for } i \in [0,I]. \nonumber
\end{eqnarray}
Moreover, definitions~\eqref{e:Jleftright} and~\eqref{e:Jdownup} are modified as
\begin{eqnarray}
\label{e:Jleftrightgro}
& \mathcal{J}^{k}_{{\rm L} (i,j)} := \eta \, \dfrac{\left(u^{k}_{(i-1,j)}-u^{k}_{(i,j)}\right)_{+}}{4 u_{\rm max} \mathcal{L}^2_k}, \quad \mathcal{J}^{k}_{{\rm R} (i,j)} := \eta \, \dfrac{\left(u^{k}_{(i+1,j)}-u^{k}_{(i,j)}\right)_{+}}{4 u_{\rm max} \mathcal{L}^2_k} \quad \text{for } (i,j)\in [1,I-1] \times [0,I], \nonumber \\\\
& \mathcal{J}^{k}_{{\rm L} (0,j)} := 0, \quad \mathcal{J}^{k}_{{\rm R} (I,j)} := 0 \quad \text{for } j \in [0,I], \nonumber
\end{eqnarray}
\begin{eqnarray}
\label{e:Jdownupgro}
& \mathcal{J}^{k}_{{\rm D} (i,j)} := \eta \, \dfrac{\left(u^{k}_{(i,j-1)}-u^{k}_{(i,j)}\right)_{+}}{4 u_{\rm max} \mathcal{L}^2_k}, \quad \mathcal{J}^{k}_{{\rm U} (i,j)} := \eta \, \dfrac{\left(u^{k}_{(i,j+1)}-u^{k}_{(i,j)}\right)_{+}}{4 u_{\rm max} \mathcal{L}^2_k} \quad \text{for } (i,j)\in [0,I] \times [1,I-1], \nonumber \\\\
& \mathcal{J}^{k}_{{\rm D} (i,0)} := 0, \quad \mathcal{J}^{k}_{{\rm U} (i,I)} := 0 \quad \text{for } i \in [0,I]. \nonumber
\end{eqnarray}
Finally, definitions~\eqref{pb}--\eqref{pq} are modified as
\begin{equation} \label{pbgro}
 \mathcal{P}_b\left(n_{{\bf i}}^{k}, u_{{\bf i}}^{k}, \mathcal{L}_{k}\right) := \tau \, \alpha_{n}\ \left(\psi(n_{{\bf i}}^{k})\right)_{+} \phi_u(u_{{\bf i}}^{k}) + \left(g_{{\bf i}}(n^{k}_{{\bf i}}, \mathcal{L}_{k})\right)_-,
 \end{equation}
 \begin{equation} \label{pdgro}
 \mathcal{P}_d\left(n_{{\bf i}}^{k}, u_{{\bf i}}^{k}, v_{{\bf i}}^{k}, \mathcal{L}_{k}\right) := \tau \, \left(\alpha_{n} \, \left(\psi(n_{{\bf i}}^{k})\right)_{-} \, \phi_u(u_{{\bf i}}^{k}) + \beta_n \, \phi_v(v_{{\bf i}}^{k})\right) + \left(g_{{\bf i}}(n^{k}_{{\bf i}}, \mathcal{L}_{k})\right)_+,
 \end{equation}
and
 \begin{equation}
 \label{pqgro}
 \mathcal{P}_q\left(n_{{\bf i}}^{k}, u_{{\bf i}}^{k}, v_{{\bf i}}^{k}, \mathcal{L}_{k} \right) := 1 - \tau \left(\alpha_{n}\ \left|\psi(n_{{\bf i}}^{k})\right| \, \phi_u(u_{{\bf i}}^{k}) + \beta_{n} \, \phi_v(v_{{\bf i}}^{k}) \right) - \left|g_{{\bf i}}(n^{k}_{{\bf i}}, \mathcal{L}_{k})\right|. 
 \end{equation}
 Here, the function $g_{{\bf i}}(n^{k}_{{\bf i}}, \mathcal{L}_{k})$ is defined via equation~\eqref{def:unifdg} in the case of uniform domain growth and via equation~\eqref{def:apicaldg} in the case of apical growth. The functions $\psi$, $\phi_u$ and $\phi_v$ satisfy assumptions~\eqref{ass:psi} and~\eqref{ass:phi}, and we assume $\tau$ and $\mathcal{L}_{k}$ to be such that that $0 < \mathcal{P}_h < 1$ for all $h \in \{b,d,q \}$. 
 
 \subsection{Corresponding continuum model}
\label{continuum_growing}
Similarly to the case of static domains, letting the time-step $\tau \to 0$ and the space-step $\chi \to 0$ in such a way that conditions~\eqref{ass:quottozero} are met, it is possible to formally show (see Appendix~\ref{app:derivation}) that the deterministic continuum counterpart of the stochastic IB model on growing domains is given by the following PDE for the cell density $n(t,\hat{{\bf x}})$
\begin{equation}
\label{eq:PDEn_growing}
\partial_t n - \nabla_{\hat {\bf x}} \cdot \left(\dfrac{D_{n}}{\mathcal{L}^2} \, \nabla_{\hat {\bf x}} n - \dfrac{C_{n}}{\mathcal{L}^2} \, n \, \nabla_{\hat {\bf x}} u \right) = \Big(\alpha_{n}\ \psi(n) \ \phi_u(u) - \beta_n \ \phi_v(v) \Big)\ n + G(\hat {\bf x}, n,\mathcal{L}), \quad (t,\hat{{\bf x}}) \in \mathbb{R}^*_+ \times (0,1)^d
\end{equation}
subject to zero-flux boundary conditions, with either
\begin{equation}
\label{def:unifdgcont}
G(\hat {\bf x}, w,\mathcal{L}) \equiv G(w,\mathcal{L}) := - d \, w \, \dfrac{1}{\mathcal{L}} \, \dfrac{{\rm d}\mathcal{L}}{{\rm d}t},
\end{equation}
in the case of uniform domain growth, or 
\begin{equation}
\label{def:apicalcont}
G(\hat {\bf x}, w,\mathcal{L}) := {\hat {\bf x}} \cdot \nabla_{\hat {\bf x}} w \, \dfrac{1}{\mathcal{L}} \, \dfrac{{\rm d}\mathcal{L}}{{\rm d}t},
\end{equation}
in the case of apical growth. Here, $D_{n} \in \mathbb{R}^*_+$ in definitions~\eqref{ass:quottozero} is the rescaled diffusion coefficient (\emph{i.e.}, the rescaled motility) of the cells, while $C_{n} \in \mathbb{R}_+$ in definitions~\eqref{ass:quottozero} represents the chemotactic sensitivity of cells to the activator. In equation~\eqref{eq:PDEn_growing}, the concentration of the activator $u(t,\hat{{\bf x}})$ and the concentration of the inhibitor $v(t,\hat{{\bf x}})$ are governed by the continuum counterpart of the difference equations~\eqref{eq:discuvgro} complemented with zero-flux boundary conditions, that is, the following system of PDEs subject to zero-flux boundary conditions
\begin{equation}
\label{eq:PDEuv_growing}
\begin{cases}
\displaystyle{\partial_t u - \dfrac{D_{u}}{\mathcal{L}^2} \, \Delta_{\hat {\bf x}} u = P(u,v) + G(\hat {\bf x}, u,\mathcal{L}),}
\\\\
\displaystyle{\partial_t v - \dfrac{D_{v}}{\mathcal{L}^2} \, \Delta_{\hat {\bf x}} v = Q(u,v) + G(\hat {\bf x}, v,\mathcal{L}),}
\end{cases}
\quad (t,\hat{{\bf x}}) \in \mathbb{R}^*_+ \times (0,1)^d
\end{equation}
which can be formally obtained by letting $\tau \to 0$ and $\chi \to 0$ in system~\eqref{eq:discuvgro}.

\subsection{Numerical simulations}
\label{numsim2}
In this section, we carry out a systematic quantitative comparison between the results of numerical simulations of the hybrid model presented in Section~\ref{hybrid_growing} and numerical solutions of the corresponding continuum model given in Section~\ref{continuum_growing}, both in one and in two spatial dimensions. All simulations are performed in {\sc{Matlab}} and the final time of simulations is chosen such that the essential features of the pattern formation process are evident. 

\subsubsection{Summary of the set-up of numerical simulations}
We define the functions $P$, $Q$, $\psi$, $\phi_u$ and $\phi_v$ as in the case of static domains. In more detail, $P$ and $Q$ are provided in definitions~\eqref{schnackenbergA}, $\psi$ is defined via equation~\eqref{def:psi}, and $\phi_u$ and $\phi_v$ are given via either definitions~\eqref{def:phi} or definitions~\eqref{def:phie1}.\\

In all simulations, we let the initial spatial distributions of morphogens and cells be the numerical steady state distributions obtained in the case of static domains with $\ell:=1$, and we assume the domain to slowly grow linearly over time, that is,
\begin{equation}
\mathcal{L}(t) :=1+ 0.01 \, t. \label{defineL}
\end{equation}
Given the values of the parameters chosen to carry out numerical simulations of the IB model, we describe $D_{n}$ and $C_{n}$ via the definitions~\eqref{ass:quottozeronum} so that conditions~\eqref{ass:quottozero} are met. A complete description of the set-up of numerical simulations is given in Appendix~\ref{app:gro}.

\subsubsection{Main results of numerical simulations}

\paragraph{Dynamics of the morphogens} The plots in the top rows of Figures~\ref{fig1Dproliferationgrow} and~\ref{fig1Dchemotaxisgrow} and in the Supplementary Figure~\ref{fig2Dprepatternbothgrowing} summarise the dynamics of the continuum concentrations of morphogens $u(t,\hat {\bf x})$ and $v(t,\hat {\bf x})$ obtained by solving numerically the system of PDEs~\eqref{eq:PDEuv_growing} subject to zero-flux boundary conditions and with $G(\hat {\bf x}, u,\mathcal{L})$ and $G(\hat {\bf x}, v,\mathcal{L})$ defined via equation~\eqref{def:unifdgcont}, while the plots in the top rows of Figures~\ref{fig1DproliferationgrowTip}~and~\ref{fig1DchemotaxisgrowTip} and in the Supplementary Figure~\ref{fig2DprepatternbothgrowingTip} refer to the case where $G(\hat {\bf x}, u,\mathcal{L})$ and $G(\hat {\bf x}, v,\mathcal{L})$ are defined via equation~\eqref{def:apicalcont}. Identical results hold for the discrete morphogen concentrations $u_{\bf i}^k$ and $v_{\bf i}^k$ obtained by solving the system of difference equations~\eqref{eq:discuvgro} (results not shown). These plots demonstrate that, when the spatial domain grows over time, a dynamical rescaling and repetition of the Turing pre-patterns observed in the case of static domains occurs---{i.e.}, spots of high concentration repeatedly split symmetrically. In the case of uniform domain growth, such a self-similar process occurs throughout the whole domain, while in the case of apical growth the process takes place toward the growing edge of the domain.

\paragraph{Dynamics of the cells}
The plots in the bottom row of Figure~\ref{fig1Dproliferationgrow} and the plots in Figure~\ref{fig2Dproliferationgrow} summarise the dynamics of the cell density in the case where there is no chemotaxis, chemically-controlled cell proliferation occurs---{i.e.}, when $\eta=0$, $C_n=0$, and the functions $\phi_u$ and $\phi_v$ are given by definitions~\eqref{def:phi}---and the functions $g_{{\bf i}}(n^{k}_{{\bf i}}, \mathcal{L}_{k})$ and $G(\hat {\bf x}, n,\mathcal{L})$ are defined via equations~\eqref{def:unifdg} and~\eqref{def:unifdgcont}, respectively. On the other hand, the plots in the bottom row of Figure~\ref{fig1DproliferationgrowTip} and the plots in Figure~\ref{fig2DproliferationgrowTip} refer to the case where the functions $g_{{\bf i}}(n^{k}_{{\bf i}}, \mathcal{L}_{k})$ and $G(\hat {\bf x}, n,\mathcal{L})$ are defined via equations~\eqref{def:apicaldg} and~\eqref{def:apicalcont}. These plots indicate that, when the spatial domain grows over time, spots of high cell density stretch either throughout the domain (uniform growth) or at the growing edge (apical growth) before splitting. This process causes cell patterns to rescale and repeat across the domain at a smaller scale. These plots demonstrate also that there is a good quantitative match between the discrete cell density $n^k_{\bf i}$ given by equation~\eqref{e:n}, with $N^k_{\bf i}$ obtained through computational simulations of the IB model, and the continuum cell density $n(t,\hat {\bf x})$ obtained by solving numerically the PDE~\eqref{eq:PDEn_growing} subject to zero-flux boundary conditions and complemented with either equation~\eqref{def:unifdgcont} or equation~\eqref{def:apicalcont}, both in one and in two spatial dimensions. 

Analogous considerations apply to the case where cell proliferation is not regulated by the morphogens and chemotactic movement of the cells up the concentration gradient of the activator occurs---{i.e.}, when the functions $\phi_u$ and $\phi_v$ are described through the definitions~\eqref{def:phie1}, $\eta>0$ and $C_n>0$---see the plots in the bottom row of Figure~\ref{fig1Dchemotaxisgrow} along with the plots in Figure~\ref{fig2Dchemotaxisgrow} and the plots in the bottom row of Figure~\ref{fig1DchemotaxisgrowTip} along with the plots in Figure~\ref{fig2DchemotaxisgrowTip}.

 \begin{figure}\centering
\includegraphics[width=\textwidth]{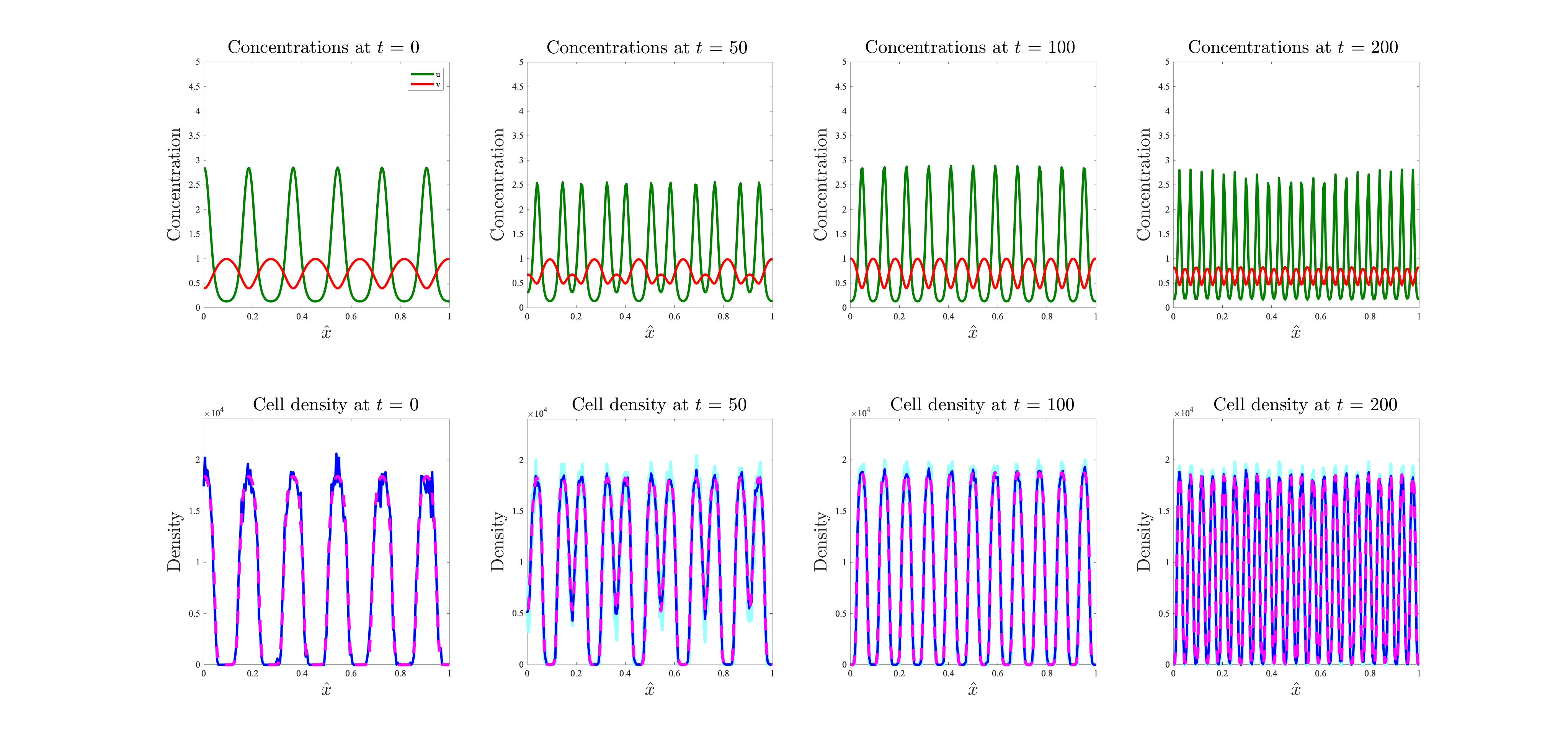}
\caption{{Results of numerical simulations on a one-dimensional uniformly growing domain in the presence of chemically-controlled cell proliferation}. {(Top row)} Plots of the concentrations of morphogens at four consecutive time instants. The green lines highlight the concentration of activator $u(t,\hat x)$ and the red lines highlight the concentration of inhibitor $v(t,\hat x)$ obtained by solving numerically the system of PDEs~\eqref{eq:PDEuv_growing} for $d=1$ subject to zero-flux boundary conditions, and complemented with the definitions~\eqref{schnackenbergA}, equation~\eqref{def:unifdgcont} and equation~\eqref{defineL}. {(Bottom row)} Comparison between the discrete cell density $n^k_i$ obtained by averaging the results of computational simulations of the IB model (solid dark blue lines) and the continuum cell density $n(t,\hat x)$ obtained by solving numerically the PDE~\eqref{eq:PDEn_growing} for $d=1$ subject to zero-flux boundary conditions and complemented with equations~\eqref{def:unifdgcont} and~\eqref{defineL} (pink dashed lines), at four consecutive time instants. Here, $\eta=0$, $C_n=0$, and the functions $\phi_u$ and $\phi_v$ are given by definitions~\eqref{def:phi}. We additionally set the initial cell density $n_{i}^{0}= 10^4$ for all $i$. The results from the IB model correspond to the average over five realisations of the underlying branching random walk, with the results from each realisation plotted in pale blue to demonstrate the robustness of the results obtained. A complete description of the set-up of numerical simulations is given in Appendix~\ref{app:gro}.\label{fig1Dproliferationgrow}}
\end{figure}

 \begin{figure}\centering
\includegraphics[width=\textwidth]{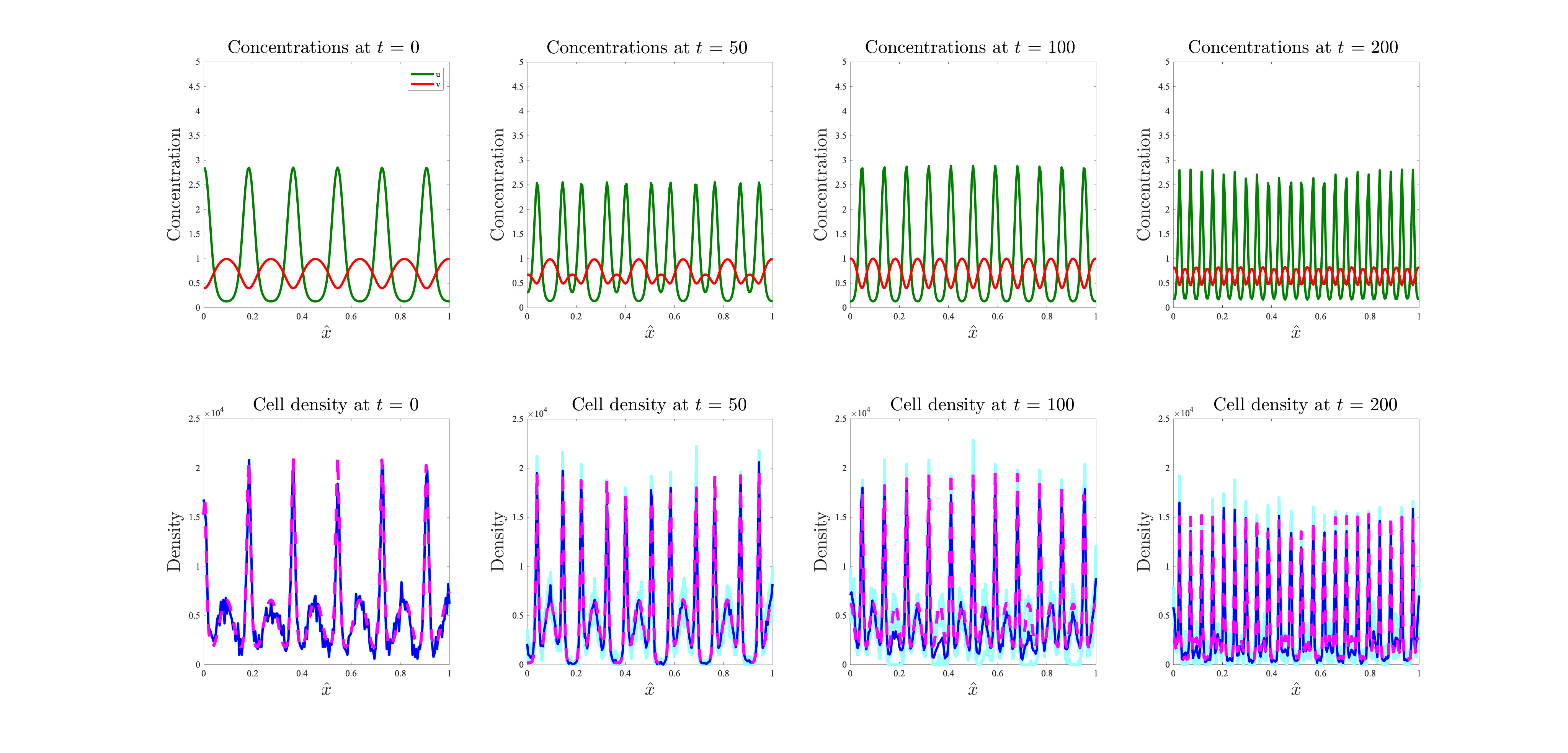}
\caption{{Results of numerical simulations on a one-dimensional uniformly growing domain in the presence of chemotaxis}. {(Top row)} Plots of the concentrations of morphogens at four consecutive time instants. The green lines highlight the concentration of activator $u(t,\hat x)$ and the red lines highlight the concentration of inhibitor $v(t,\hat x)$ obtained by solving numerically the system of PDEs~\eqref{eq:PDEuv_growing} for $d=1$ subject to zero-flux boundary conditions, and complemented with the definitions~\eqref{schnackenbergA}, equation~\eqref{def:unifdgcont} and equation~\eqref{defineL}. {(Bottom row)} Comparison between the discrete cell density $n^k_i$ obtained by averaging the results of computational simulations of the IB model (solid dark blue lines) and the continuum cell density $n(t,\hat x)$ obtained by solving numerically the PDE~\eqref{eq:PDEn_growing} for $d=1$ subject to zero-flux boundary conditions and complemented with equations~\eqref{def:unifdgcont} and~\eqref{defineL} (pink dashed lines), at four consecutive time instants. Here, $\eta>0$, $C_n>0$, and the functions $\phi_u$ and $\phi_v$ are described through the definitions~\eqref{def:phie1}. We additionally set the initial cell density $n_{i}^{0}= 10^4$ for all $i$. The results from the IB model correspond to the average over five realisations of the underlying branching random walk, with the results from each realisation plotted in pale blue to demonstrate the robustness of the results obtained. A complete description of the set-up of numerical simulations is given in Appendix~\ref{app:gro}.\label{fig1Dchemotaxisgrow}}
\end{figure}

 \begin{figure}\centering
\includegraphics[width=\textwidth]{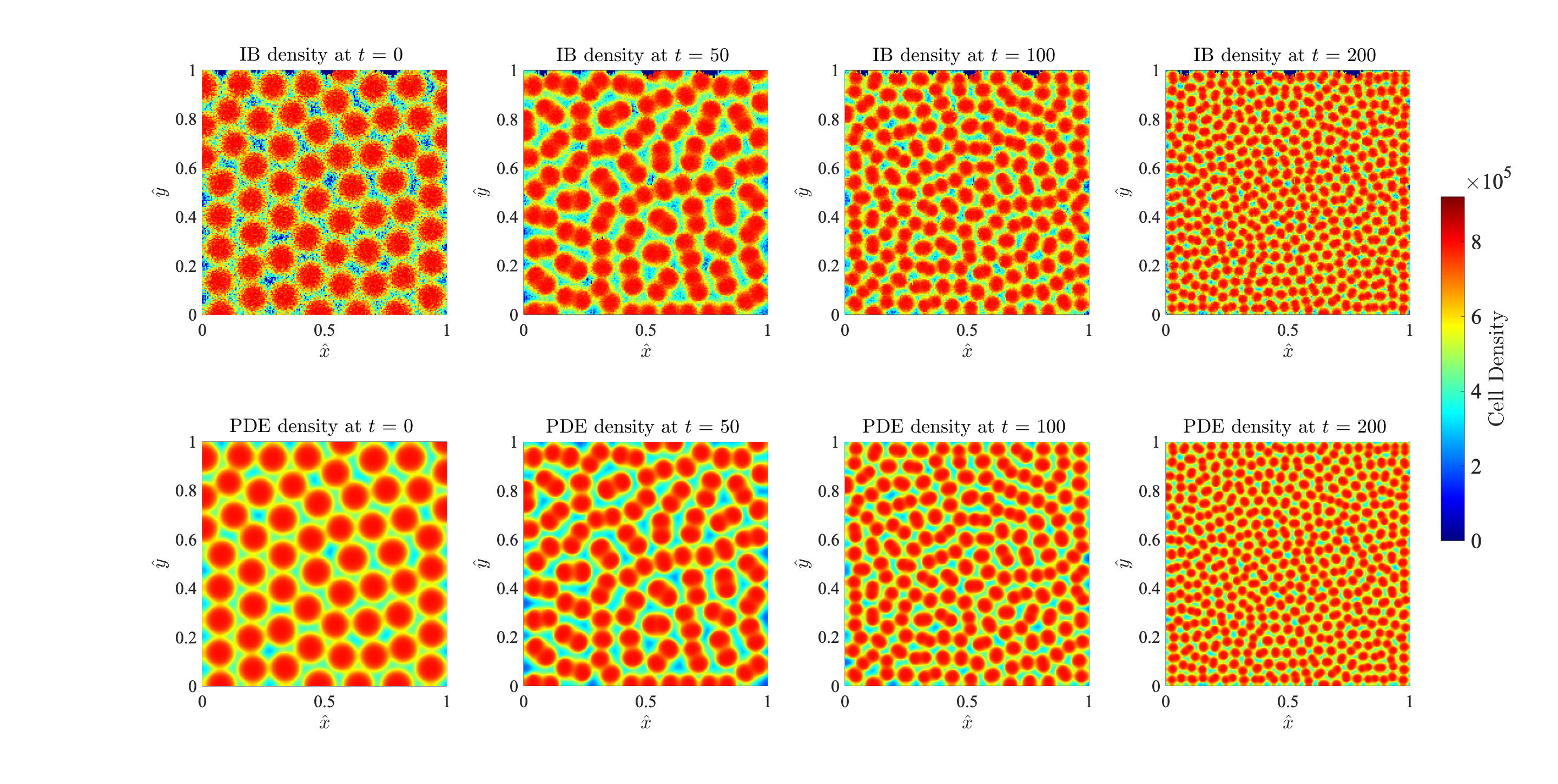}
\caption{{Results of numerical simulations on a two-dimensional uniformly growing domain in the presence of chemically-controlled cell proliferation}. Comparison between the discrete cell density $n^k_{\bf i}$ obtained by averaging the results of computational simulations of the IB model (top row) and the continuum cell density $n(t,\hat {\bf x})$ obtained by solving numerically the PDE~\eqref{eq:PDEn_growing} for $d=2$ subject to zero-flux boundary conditions and complemented with equations~\eqref{def:unifdgcont} and~\eqref{defineL} (bottom row), at four consecutive time instants. Here, $\eta=0$, $C_n=0$, and the functions $\phi_u$ and $\phi_v$ are given by definitions~\eqref{def:phi}. We additionally set the initial cell density $n_{\bf i}^{0}=4\times10^5$ for all ${\bf i}$. The results from the IB model correspond to the average over five realisations of the underlying branching random walk. The plots of the corresponding morphogen concentrations are displayed in the Supplementary Figure~\ref{fig2Dprepatternbothgrowing}. A complete description of the set-up of numerical simulations is given in Appendix~\ref{app:gro}.\label{fig2Dproliferationgrow}}
\end{figure}

 \begin{figure}\centering
\includegraphics[width=\textwidth]{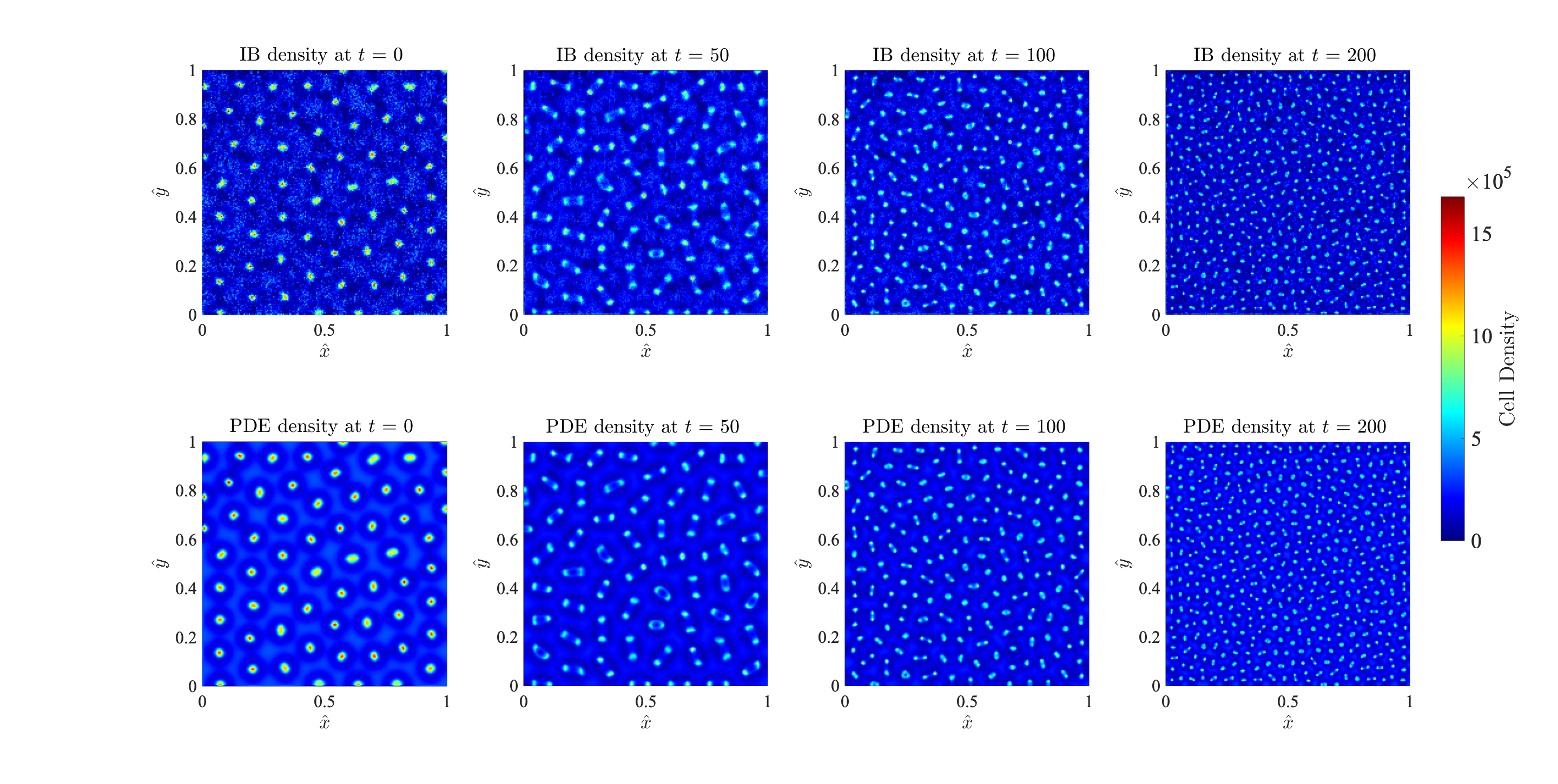}
\caption{{Results of numerical simulations on a two-dimensional uniformly growing domain in the presence of chemically-controlled cell proliferation}. Comparison between the discrete cell density $n^k_{\bf i}$ obtained by averaging the results of computational simulations of the IB model (top row) and the continuum cell density $n(t,\hat {\bf x})$ obtained by solving numerically the PDE~\eqref{eq:PDEn_growing} for $d=2$ subject to zero-flux boundary conditions and complemented with equations~\eqref{def:unifdgcont} and~\eqref{defineL} (bottom row), at four consecutive time instants. Here, $\eta>0$, $C_n>0$, and the functions $\phi_u$ and $\phi_v$ are described through the definitions~\eqref{def:phie1}. We additionally set the initial cell density $n_{\bf i}^{0}=4\times10^5$ for all ${\bf i}$. The results from the IB model correspond to the average over five realisations of the underlying branching random walk. The plots of the corresponding morphogen concentrations are displayed in the Supplementary Figure~\ref{fig2Dprepatternbothgrowing}. A complete description of the set-up of numerical simulations is given in Appendix~\ref{app:gro}.\label{fig2Dchemotaxisgrow}}
\end{figure}

 \begin{figure}\centering
\includegraphics[width=\textwidth]{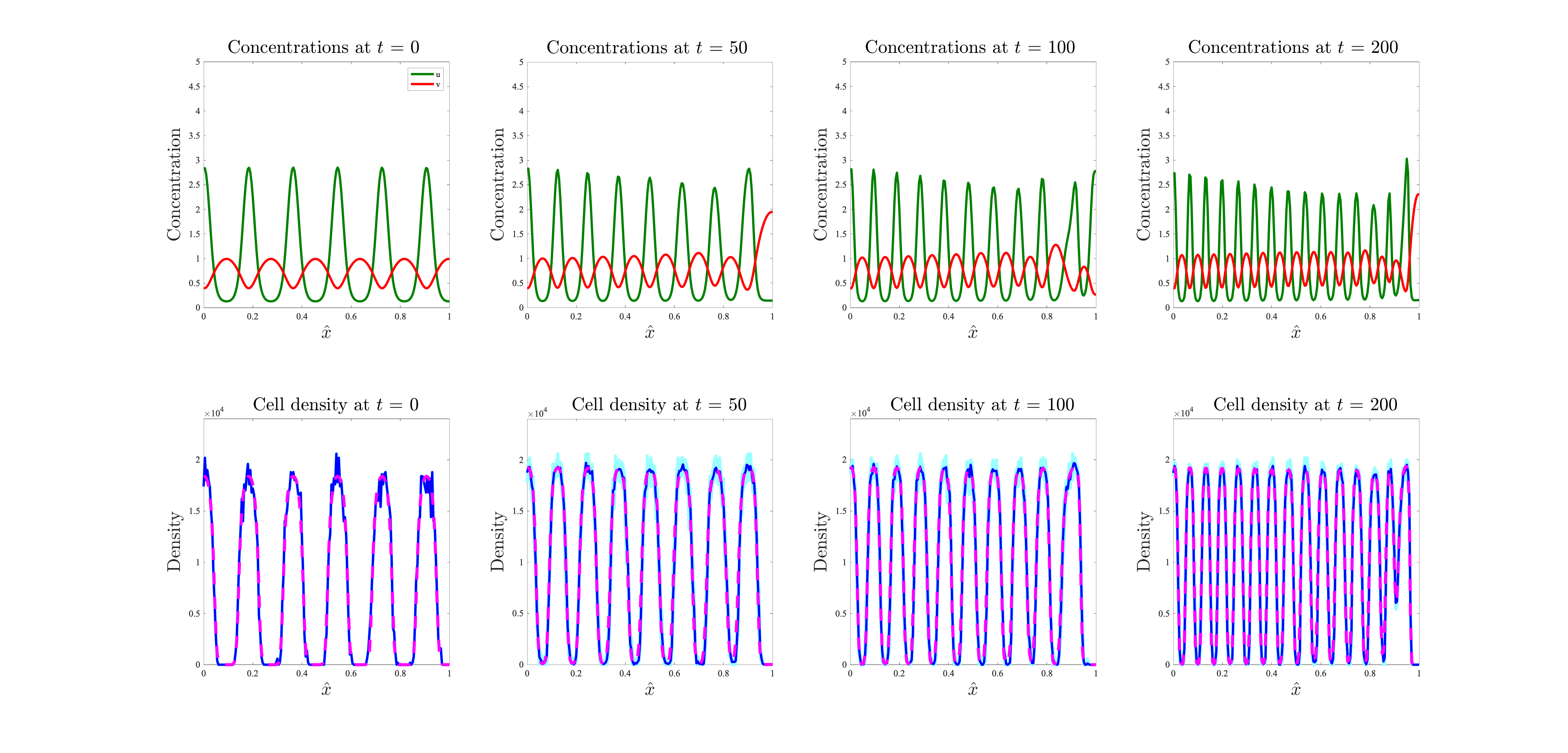}
\caption{{Results of numerical simulations on a one-dimensional apically growing domain in the presence of chemically-controlled cell proliferation}. {(Top row)} Plots of the concentrations of morphogens at four consecutive time instants. The green lines highlight the concentration of activator $u(t,\hat x)$ and the red lines highlight the concentration of inhibitor $v(t,\hat x)$ obtained by solving numerically the system of PDEs~\eqref{eq:PDEuv_growing} for $d=1$ subject to zero-flux boundary conditions, complemented with the definitions~\eqref{schnackenbergA}, equation~\eqref{def:apicalcont} and equation~\eqref{defineL}. {(Bottom row)} Comparison between the discrete cell density $n^k_i$ obtained by averaging the results of computational simulations of the IB model (solid dark blue lines) and the continuum cell density $n(t,\hat x)$ obtained by solving numerically the PDE~\eqref{eq:PDEn_growing} for $d=1$ subject to zero-flux boundary conditions and complemented with equations~\eqref{def:apicalcont} and~\eqref{defineL} (pink dashed lines), at four consecutive time instants. Here, $\eta=0$, $C_n=0$, and the functions $\phi_u$ and $\phi_v$ are given by definitions~\eqref{def:phi}. We additionally set the initial cell density $n_{i}^{0}= 10^4$ for all $i$. The results from the IB model correspond to the average over five realisations of the underlying branching random walk, with the results from each realisation plotted in pale blue to demonstrate the robustness of the results obtained. A complete description of the set-up of numerical simulations is given in Appendix~\ref{app:gro}.\label{fig1DproliferationgrowTip}}
\end{figure}

 \begin{figure}\centering
\includegraphics[width=\textwidth]{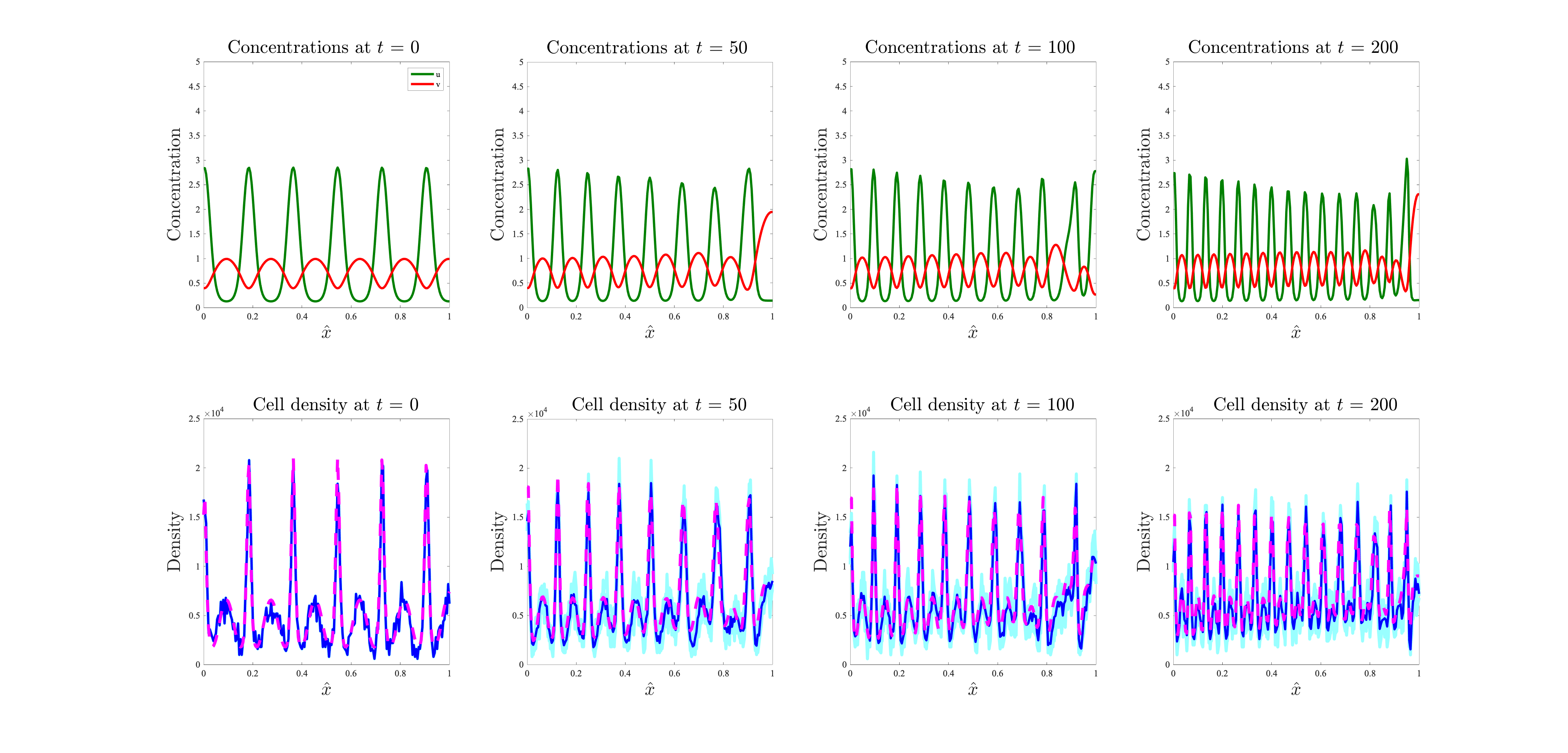}
\caption{{Results of numerical simulations on a one-dimensional apically growing domain in the presence of chemotaxis}. {(Top row)} Plots of the concentrations of morphogens at four consecutive time instants. The green lines highlight the concentration of activator $u(t,\hat x)$ and the red lines highlight the concentration of inhibitor $v(t,\hat x)$ obtained by solving numerically the system of PDEs~\eqref{eq:PDEuv_growing} for $d=1$ subject to zero-flux boundary conditions, complemented with the definitions~\eqref{schnackenbergA}, equation~\eqref{def:apicalcont} and equation~\eqref{defineL}. {(Bottom row)} Comparison between the discrete cell density $n^k_i$ obtained by averaging the results of computational simulations of the IB model (solid dark blue lines) and the continuum cell density $n(t,\hat x)$ obtained by solving numerically the PDE~\eqref{eq:PDEn_growing} for $d=1$ subject to zero-flux boundary conditions and complemented with equations~\eqref{def:apicalcont} and~\eqref{defineL} (pink dashed lines), at four consecutive time instants. Here, $\eta>0$, $C_n>0$, and the functions $\phi_u$ and $\phi_v$ are described through the definitions~\eqref{def:phie1}. We additionally set the initial cell density $n_{i}^{0}= 10^4$ for all $i$. The results from the IB model correspond to the average over five realisations of the underlying branching random walk, with the results from each realisation plotted in pale blue to demonstrate the robustness of the results obtained. A complete description of the set-up of numerical simulations is given in Appendix~\ref{app:gro}.\label{fig1DchemotaxisgrowTip}}
\end{figure}

 \begin{figure}\centering
\includegraphics[width=\textwidth]{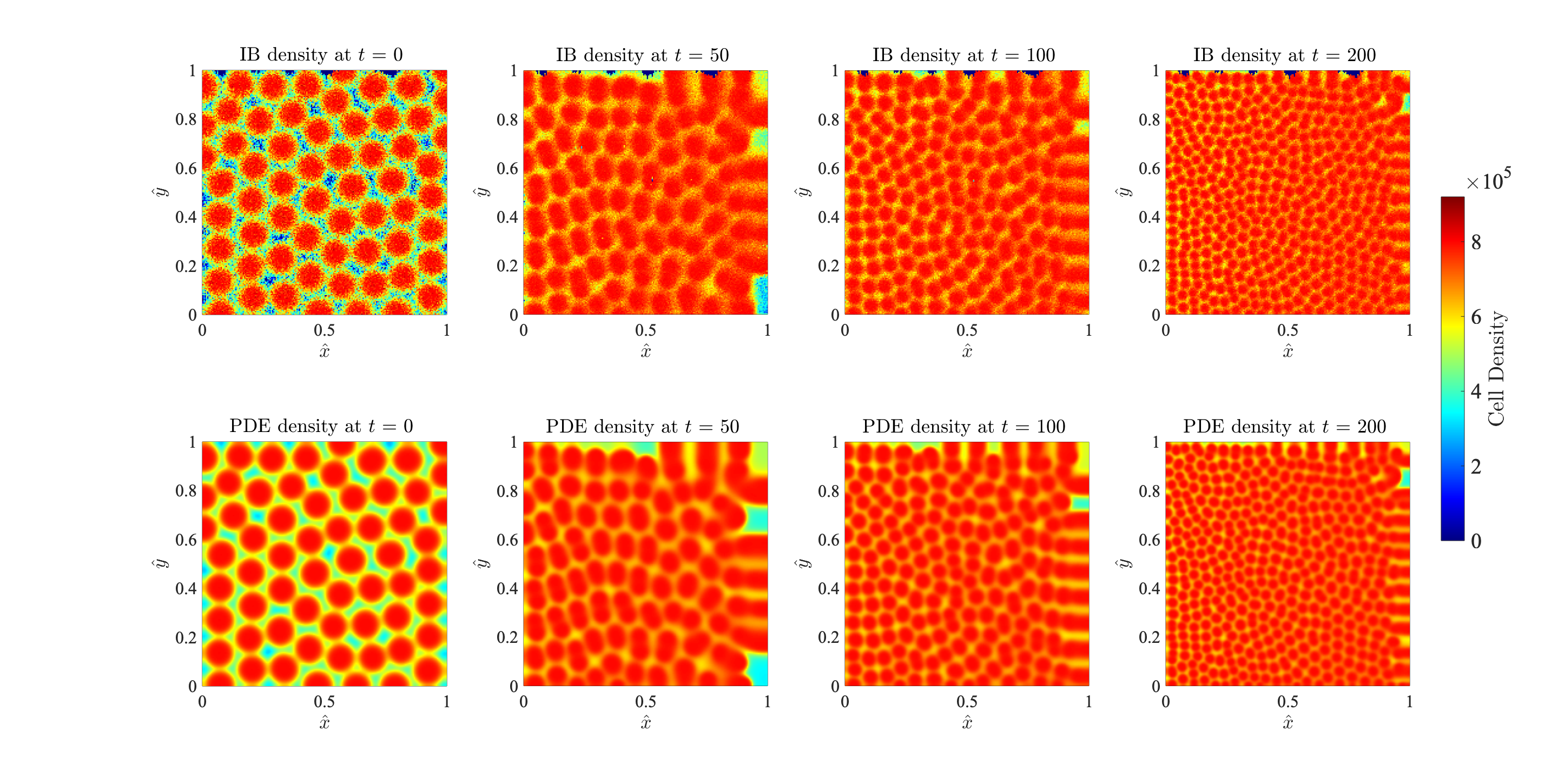}
\caption{{Results of numerical simulations on a two-dimensional apically growing domain in the presence of chemically-controlled cell proliferation}. Comparison between the discrete cell density $n^k_{\bf i}$ obtained by averaging the results of computational simulations of the IB model (top row) and the continuum cell density $n(t,\hat {\bf x})$ obtained by solving numerically the PDE~\eqref{eq:PDEn_growing} for $d=2$ subject to zero-flux boundary conditions and complemented with equations~\eqref{def:apicalcont} and~\eqref{defineL} (bottom row), at four consecutive time instants. Here, $\eta=0$, $C_n=0$, and the functions $\phi_u$ and $\phi_v$ are given by definitions~\eqref{def:phi}. We additionally set the initial cell density $n_{\bf i}^{0}=4\times10^5$ for all ${\bf i}$. The results from the IB model correspond to the average over five realisations of the underlying branching random walk. The plots of the corresponding morphogen concentrations are displayed in the Supplementary Figure~\ref{fig2DprepatternbothgrowingTip}. A complete description of the set-up of numerical simulations is given in Appendix~\ref{app:gro}.\label{fig2DproliferationgrowTip}}
\end{figure}

 \begin{figure}\centering
\includegraphics[width=\textwidth]{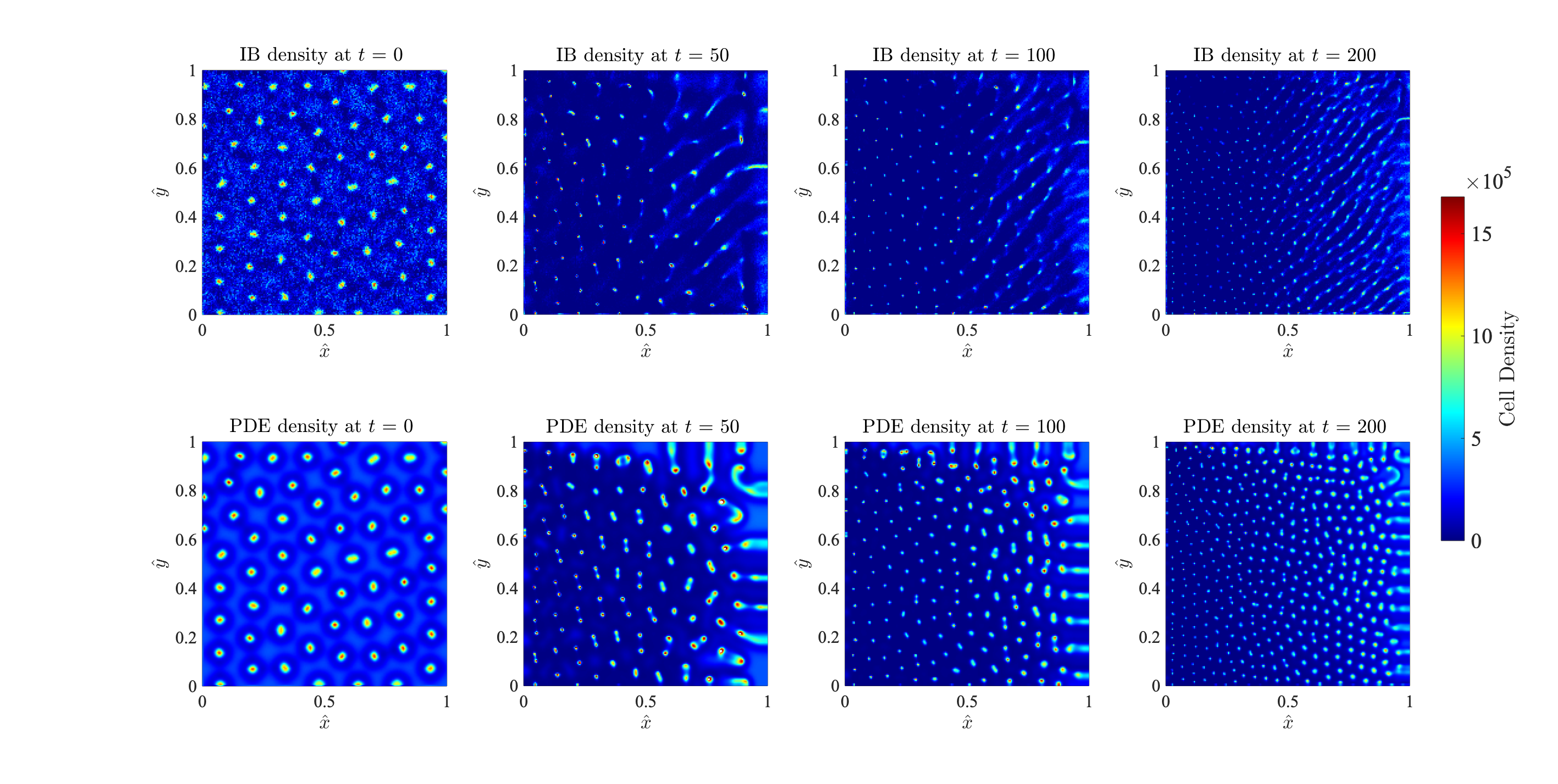}
\caption{{Results of numerical simulations on a two-dimensional apically growing domain in the presence of chemically-controlled cell proliferation}. Comparison between the discrete cell density $n^k_{\bf i}$ obtained by averaging the results of computational simulations of the IB model (top row) and the continuum cell density $n(t,\hat {\bf x})$ obtained by solving numerically the PDE~\eqref{eq:PDEn_growing} for $d=2$ subject to zero-flux boundary conditions and complemented with equations~\eqref{def:apicalcont} and~\eqref{defineL} (bottom row), at four consecutive time instants. Here, $\eta>0$, $C_n>0$, and the functions $\phi_u$ and $\phi_v$ are described through the definitions~\eqref{def:phie1}. We additionally set the initial cell density $n_{\bf i}^{0}=4\times10^5$ for all ${\bf i}$. The results from the IB model correspond to the average over five realisations of the underlying branching random walk. The plots of the corresponding morphogen concentrations are displayed in the Supplementary Figure~\ref{fig2DprepatternbothgrowingTip}. A complete description of the set-up of numerical simulations is given in Appendix~\ref{app:gro}.\label{fig2DchemotaxisgrowTip}}
\end{figure}

\section{Conclusions and research perspectives}
\label{sect:conclusion}
\subsection{Conclusions} 
We have developed a hybrid discrete-continuum modelling framework that can be used to describe the formation of cellular patterns, specifically focusing on the Turing mechanism as the driving force behind the patterns. We used reaction-diffusion systems to describe the evolution of morphogens, which dictate the action of cells, while cell dynamics were described by stochastic IB models. We formally derived the deterministic continuum counterparts of the IB models, which were formulated as PDEs for the cell density, and compared the two modelling approaches through numerical simulations both in the case of stationary spatial domains and in the case of two types of growing domains, corresponding to uniform and apical growth. Numerical simulations demonstrated that in the case of sufficiently large cell numbers there was an excellent quantitative match between the spatial patterns produced by the stochastic IB model and its deterministic continuum counterpart. Moreover, in the case of static domains, we also presented the results of numerical simulations showing that possible differences between the spatial patterns produced by the two modelling approaches could emerge in the regime of sufficiently low cell numbers. In fact, lower cell numbers correlated with both lower regularity of the cell density and demographic stochasticity, which may cause a reduction in the quality of the approximations employed in the formal derivation of the deterministic continuum model from the stochastic IB model. Hence, having both types of models available allows one to use IB models in the regime of low cells numbers---{i.e.}, when stochastic effects associated with small cell population levels, which cannot be captured by PDE models, are particularly relevant---and then turn to their less computationally expensive PDE counterparts when large cell numbers need to be considered---i.e., when stochastic effects associated with small cell population levels are negligible. \\

\subsection{Research perspectives}
There are a number of additional elements that would be relevant to incorporate into the modelling framework presented here in order to further broaden its spectrum of applications. 

For instance, as was recognised by Turing himself, exogenous diffusing chemicals are not the only vehicle of coordination between cells. In particular, it is known that long range cell-cell interactions can be mediated by signal proteins produced by the cells themselves and also by mechanical forces between cells and components of the cellular microenvironment. For example, vascular endothelial growth factor signalling has been shown to control neural crest cell migration~\cite{mclennan2012multiscale,mclennan2015neural,mclennan2015vegf}, and mechanical interactions between cells and the extra cellular matrix can control cell aggregation~\cite{murray1983mechanical}. Moreover, cellular patterning leading to the emergence of spatial structures often requires the interplay between non-diffusible species, transcription factors and cell signalling---\emph{viz.} the process underlying digit formation in tetrapods~\cite{schweisguth2019self}. In this regard, it would be interesting to extend the modelling framework by allowing the cells to consume and/or produce chemicals required for successful coordination of their actions~\cite{tweedy2016self}, and by incorporating more complex cellular processes such as anoikis~\cite{galle2005modeling,galle2006individual} and cell deformation~\cite{drasdo2007role,neilson2011chemotaxis}. In the situation where local production and/or consumption of the chemicals by the cells occurs, for particular cases such as those considered in~\cite{bubba2020discrete}, we would still expect it to be possible to derive an effective deterministic continuum limit of the IB model for the dynamics of the cells through formal procedures analogous to the one used here. However, there could also be cases in which PDE models derived using similar formal procedures might not be able to faithfully reproduce the dynamics of the branching random walk underlying the IB model, due to the interplay between stochastic effects and nonlinear dynamical interactions between the cells and the chemicals.

To date, only few biological systems have been demonstrated to satisfy the necessary conditions required for the formation of Turing pre-patterns via reaction-diffusion systems. Since mathematical models formulated as scalar integro-differential equations, whereby the formation of Turing-like patterns is governed by suitable integral kernels, have proven capable of faithfully reproduce a variety of pigmentation patterns in fish~\cite{kondo2012turing,kondo2017updated}, it would also be interesting to explore possible ways of integrating such alternative modelling strategies into our framework. 

\section*{Acknowledgments }
MAJC gratefully acknowledges support of EPSRC Grant No. EP/N014642/1 (EPSRC Centre for Multiscale Soft Tissue Mechanics--With Application to Heart \& Cancer).
\section*{Conflict of interest}
All authors declare no conflicts of interest in this paper.

\providecommand{\href}[2]{#2}
\providecommand{\arxiv}[1]{\href{http://arxiv.org/abs/#1}{arXiv:#1}}
\providecommand{\url}[1]{\texttt{#1}}
\providecommand{\urlprefix}{URL }

\appendix
\setcounter{figure}{0}
\renewcommand\thefigure{\thesection\arabic{figure}} 
\section*{Appendix}
\section{Formal derivation of the deterministic continuum model on growing domains}
\label{app:derivation}
We carry out a formal derivation of the deterministic continuum model given by the PDE~\eqref{ass:quottozero} for $d=2$. Similar methods can be used in the case where $d=1$. 

When cell dynamics are governed by the rules described in Section~\ref{describeIB} and Section~\ref{describeIBgro}, considering $(i,j) \in [1,I-1] \times [1,I-1]$, the mass balance principle gives 
\begin{eqnarray}
\label{app:disc1}
n_{(i,j)}^{k+1}&=&n_{(i,j)}^{k}+\frac{\theta}{4 \mathcal{L}^2_k}\Big[n_{(i+1,j)}^{k}+n_{(i-1,j)}^{k}+n_{(i,j+1)}^{k}+n_{(i,j-1)}^{k}-4n_{(i,j)}^{k}\Big]\nonumber\\
&&+\frac{\eta}{4\ u_{\rm max} \mathcal{L}^2_k}\Big[\Big(u_{(i,j)}^{k}-u_{(i-1,j)}^{k}\Big)_{+}\ n_{(i-1,j)}^{k}+\Big(u_{(i,j)}^{k}-u_{(i+1,j)}^{k}\Big)_{+}\ n_{(i+1,j)}^{k}\Big]\nonumber\\
&&+\frac{\eta}{4\ u_{\rm max} \mathcal{L}^2_k}\Big[\Big(u_{(i,j)}^{k}-u_{(i,j-1)}^{k}\Big)_{+}\ n_{(i,j-1)}^{k}+\Big(u_{(i,j)}^{k}-u_{(i,j+1)}^{k}\Big)_{+}\ n_{(i,j+1)}^{k}\Big]\nonumber\\
&&-\frac{\eta}{4\ u_{\rm max} \mathcal{L}^2_k}\Big[\Big(u_{(i-1,j)}^{k}-u_{(i,j)}^{k}\Big)_{+}+\Big(u_{(i+1,j)}^{k}-u_{(i,j)}^{k}\Big)_{+}\Big] \, n_{(i,j)}^{k} \nonumber\\
&&-\frac{\eta}{4\ u_{\rm max} \mathcal{L}^2_k}\Big[\Big(u_{(i,j-1)}^{k}-u_{(i,j)}^{k}\Big)_{+}+\Big(u_{(i,j+1)}^{k}-u_{(i,j)}^{k}\Big)_{+}\Big] \, n_{(i,j)}^{k} \nonumber\\
&& + \tau \, \Big( \alpha_{n} \psi(n_{(i,j)}^{k})\phi_{u} (u_{(i,j)}^{k})-\beta_{n}\phi_{v} (v_{(i,j)}^{k})\Big) \, n_{(i,j)}^{k} - g_{(i,j)}(n^{k}_{(i,j)}, \mathcal{L}_{k}).
\end{eqnarray}
Using the fact that the following relations hold for $\tau$ and $\chi$ sufficiently small
\begin{eqnarray}
&t_{k}\approx t,\quad t_{k+1}\approx t+\tau, \quad \hat{x}_{i}\approx \hat{x}, \quad \quad \hat{x}_{i\pm1}\approx \hat{x}\pm\chi, \quad \hat{y}_{j}\approx \hat{y}, \quad \quad \hat{y}_{j\pm1}\approx \hat{y}\pm\chi\nonumber\\
&n_{(i,j)}^{k}\approx n(t,\hat x, \hat y),\quad n_{(i,j)}^{k+1}\approx n(t+\tau,\hat x, \hat y), \quad n_{(i\pm1,j)}^{k}\approx n(t,\hat{x}\pm\chi,\hat{y}),\quad n_{(i,j\pm1)}^{k}\approx n(t,\hat{x},\hat{y}\pm\chi), \nonumber\\
&u_{(i,j)}^{k}\approx u(t,\hat x, \hat y),\quad u_{(i,j)}^{k+1}\approx u(t+\tau,\hat x, \hat y), \quad u_{(i\pm1,j)}^{k}\approx u(t,\hat{x}\pm\chi,\hat{y}),\quad u_{(i,j\pm1)}^{k}\approx u(t,\hat{x},\hat{y}\pm\chi), \nonumber\\
&v_{(i,j)}^{k}\approx v(t,\hat x, \hat y),\quad v_{(i,j)}^{k+1}\approx v(t+\tau,\hat x, \hat y), \quad v_{(i\pm1,j)}^{k}\approx v(t,\hat{x}\pm\chi,\hat{y}),\quad v_{(i,j\pm1)}^{k}\approx v(t,\hat{x},\hat{y}\pm\chi), \nonumber\\
& \mathcal{L}_{k}\approx \mathcal{L}(t), \quad \mathcal{L}_{k+1}\approx \mathcal{L}(t+\tau), \nonumber
\end{eqnarray}
the balance equation~\eqref{app:disc1} can be formally rewritten in the approximate form
\begin{eqnarray}
\label{app:disc2}
n(t+\tau,\hat x, \hat y)&=&n+\frac{\theta}{4 \mathcal{L}^2}\Big[n(t,\hat{x}+\chi,\hat{y})+n(t,\hat{x}-\chi,\hat{y})+n(t,\hat{x},\hat{y}+\chi)+n(t,\hat{x},\hat{y}-\chi)-4n\Big]\nonumber\\
&&+\frac{\eta}{4\ u_{\rm max} \mathcal{L}^2}\Big[\Big(u-u(t,\hat{x}-\chi,\hat{y})\Big)_{+}\ n(t,\hat{x}-\chi,\hat{y})+\Big(u-u(t,\hat{x}+\chi,\hat{y})\Big)_{+}\ n(t,\hat{x}+\chi,\hat{y})\Big]\nonumber\\
&&+\frac{\eta}{4\ u_{\rm max} \mathcal{L}^2}\Big[\Big(u-u(t,\hat{x},\hat{y}-\chi)\Big)_{+}\ n(t,\hat{x},\hat{y}-\chi)+\Big(u-u(t,\hat{x},\hat{y}+\chi)\Big)_{+}\ n(t,\hat{x},\hat{y}+\chi)\Big]\nonumber\\
&&-\frac{\eta}{4\ u_{\rm max} \mathcal{L}^2}\Big[\Big(u(t,\hat{x}-\chi,\hat{y})-u\Big)_{+}+\Big(u(t,\hat{x}+\chi,\hat{y})-u\Big)_{+}\Big] \, n \nonumber\\
&&-\frac{\eta}{4\ u_{\rm max} \mathcal{L}^2}\Big[\Big(u(t,\hat{x},\hat{y}-\chi)-u\Big)_{+}+\Big(u(t,\hat{x},\hat{y}+\chi)-u\Big)_{+}\Big] \, n \nonumber\\
&& + \tau \, \Big( \alpha_{n} \psi(n)\phi_{u} (u)-\beta_{n}\phi_{v} (v)\Big) \, n - \Gamma\left(\hat x, \hat y, n, \mathcal{L}\right),
\end{eqnarray}
with
$$
{\small
\Gamma\left(\hat x, \hat y, n, \mathcal{L}\right) := 
\begin{cases}
2 \, n \, \dfrac{\mathcal{L}(t+\tau) - \mathcal{L}(t)}{\mathcal{L}(t)}, \; &\text{if } g_{(i,j)}(n^{k}_{(i,j)}) \text{ is defined via equation~\eqref{def:unifdg},}
\\\\
\left[\dfrac{\hat x}{\chi} \, \Big(n(t,\hat{x}+\chi,\hat{y})-n\Big) + \dfrac{\hat y}{\chi} \, \Big(n(t,\hat{x},\hat{y}+\chi) - n\Big)\right] \, \dfrac{\mathcal{L}(t+\tau) - \mathcal{L}(t)}{\mathcal{L}(t)}, \; &\text{if } g_{(i,j)}(n^{k}_{(i,j)}) \text{ is defined via equation~\eqref{def:apicaldg},}
\end{cases}
}
$$
where $n \equiv n(t,\hat x, \hat y)$, $u \equiv u(t,\hat x, \hat y)$, $v \equiv v(t,\hat x, \hat y)$ and $\mathcal{L} \equiv \mathcal{L}(t)$. Dividing both sides of equation~\eqref{app:disc2} by $\tau$ gives
\begin{eqnarray}
\label{app:disc3}
\dfrac{n(t+\tau,\hat x, \hat y) - n}{\tau}&=&\frac{\theta}{4 \mathcal{L}^2 \tau}\Big[n(t,\hat{x}+\chi,\hat{y})+n(t,\hat{x}-\chi,\hat{y})+n(t,\hat{x},\hat{y}+\chi)+n(t,\hat{x},\hat{y}-\chi)-4n\Big]\nonumber\\
&&+\frac{\eta}{4\ u_{\rm max} \mathcal{L}^2 \tau}\Big[\Big(u-u(t,\hat{x}-\chi,\hat{y})\Big)_{+}\ n(t,\hat{x}-\chi,\hat{y})+\Big(u-u(t,\hat{x}+\chi,\hat{y})\Big)_{+}\ n(t,\hat{x}+\chi,\hat{y})\Big]\nonumber\\
&&+\frac{\eta}{4\ u_{\rm max} \mathcal{L}^2 \tau}\Big[\Big(u-u(t,\hat{x},\hat{y}-\chi)\Big)_{+}\ n(t,\hat{x},\hat{y}-\chi)+\Big(u-u(t,\hat{x},\hat{y}+\chi)\Big)_{+}\ n(t,\hat{x},\hat{y}+\chi)\Big]\nonumber\\
&&-\frac{\eta}{4\ u_{\rm max} \mathcal{L}^2 \tau}\Big[\Big(u(t,\hat{x}-\chi,\hat{y})-u\Big)_{+}+\Big(u(t,\hat{x}+\chi,\hat{y})-u\Big)_{+}\Big] \, n \nonumber\\
&&-\frac{\eta}{4\ u_{\rm max} \mathcal{L}^2 \tau}\Big[\Big(u(t,\hat{x},\hat{y}-\chi)-u\Big)_{+}+\Big(u(t,\hat{x},\hat{y}+\chi)-u\Big)_{+}\Big] \, n \nonumber\\
&& + \Big( \alpha_{n} \psi(n)\phi_{u} (u)-\beta_{n}\phi_{v} (v)\Big) \, n - \dfrac{1}{\tau} \, \Gamma\left(\hat x, \hat y, n, \mathcal{L}\right).
\end{eqnarray}
If $n(t, \hat x, \hat y)$ is a twice continuously differentiable function of $\hat x$ and $\hat y$ and a continuously differentiable function of $t$, $u(t, \hat x, \hat y)$ is a twice continuously differentiable function of $\hat x$ and $\hat y$, and the function $\mathcal{L}(t)$ is continuously differentiable, for $\chi$ and $\tau$ sufficiently small we can use the Taylor expansions
\begin{eqnarray}
& n(t,\hat{x}\pm\chi,\hat{y})=n\pm\chi\dfrac{\partial n}{\partial \hat{x}}+\dfrac{\chi^{2}}{2}\dfrac{\partial^{2} n}{\partial \hat{x}^{2}}+\mathcal{O}(\chi^{3}),\quad n(t,\hat{x},\hat{y}\pm\chi)=n\pm\chi\dfrac{\partial n}{\partial \hat{y}}+\dfrac{\chi^{2}}{2}\dfrac{\partial^{2} n}{\partial \hat{y}^{2}}+\mathcal{O}(\chi^{3}), \nonumber \\
& n(t+\tau, \hat x, \hat y) =n+\tau\dfrac{\partial n}{\partial t}+\mathcal{O}(\tau^{2}), \nonumber \\
& u(t,\hat{x}\pm\chi,\hat{y})=u\pm\chi\dfrac{\partial u}{\partial \hat{x}}+\dfrac{\chi^{2}}{2}\dfrac{\partial^{2} u}{\partial \hat{x}^{2}}+\mathcal{O}(\chi^{3}),\quad u(t,\hat{x},\hat{y}\pm\chi)=u\pm\chi\dfrac{\partial u}{\partial \hat{y}}+\dfrac{\chi^{2}}{2}\dfrac{\partial^{2} u}{\partial \hat{y}^{2}}+\mathcal{O}(\chi^{3}) \nonumber, \\
& \mathcal{L}(t+\tau) =\mathcal{L} + \tau \dfrac{{\rm d} \mathcal{L}}{{\rm d} t}+\mathcal{O}(\tau^{2}). \nonumber
\end{eqnarray}

Substituting into equation~\eqref{app:disc3}, using the elementary property $(a)_+ - (-a)_+ = a$ for $a \in \mathbb{R}$ and letting $\tau \to 0$ and $\chi \to 0$ in such a way that conditions~\eqref{ass:quottozero} are met, after a little algebra, as similarly done in~\cite{bubba2020discrete}, we find
\begin{eqnarray}
\label{app:disc4}
\dfrac{\partial n}{\partial t} &=&\frac{D_n}{\mathcal{L}^2}\left(\frac{\partial^{2} n}{\partial \hat{x}^{2}}+\frac{\partial^{2} n}{\partial \hat{y}^{2}}\right)+ \frac{C_n}{\mathcal{L}^2}\left[\left(\frac{\partial^{2} u}{\partial \hat{x}^{2}}+\frac{\partial^{2} u}{\partial \hat{y}^{2}}\right)n-\left(\frac{\partial u}{\partial \hat{x}}\frac{\partial n}{\partial \hat{x}}+\frac{\partial u}{\partial \hat{y}}\frac{\partial n}{\partial \hat{y}}\right)\right] \nonumber\\
&&+\Big( \alpha_{n} \psi(n)\phi_{u} (u)-\beta_{n}\phi_{v} (v)\Big) \, n - G(\hat x, \hat y, n, \mathcal{L}), \quad (t,\hat{x},\hat{y}) \in \mathbb{R}^*_+ \times (0,1) \times (0,1),
\end{eqnarray}
where $G(\hat x, \hat y, n,\mathcal{L})$ is given by equation~\eqref{def:unifdgcont} in the case where $g_{(i,j)}(n^{k}_{(i,j)})$ is defined via equation~\eqref{def:unifdg} and by equation~\eqref{def:apicalcont} in the case where $g_{(i,j)}(n^{k}_{(i,j)})$ is defined via equation~\eqref{def:apicaldg}. The PDE~\eqref{app:disc4} can be easily rewritten in the form of equation~\eqref{eq:PDEn_growing}. Moreover, zero-flux boundary conditions easily follow from the fact that [\emph{cf.} definitions~\eqref{e:diffusionLRgro}--\eqref{e:Jdownupgro}]
$$
\mathcal{T}^{k}_{{\rm L} (0,j)} := 0,\quad \mathcal{T}^{k}_{{\rm R} (I,j)} := 0, \quad \mathcal{J}^{k}_{{\rm L} (0,j)} := 0, \quad \mathcal{J}^{k}_{{\rm R} (I,j)} := 0 \quad \text{for } j \in [0,I]
$$
and
$$
\mathcal{T}^{k}_{{\rm D} (i,0)} := 0,\quad \mathcal{T}^{k}_{{\rm U} (i,I)} := 0, \quad \mathcal{J}^{k}_{{\rm D} (i,0)} := 0, \quad \mathcal{J}^{k}_{{\rm U} (i,I)} := 0 \quad \text{for } i \in [0,I].
$$ 

\begin{remark}
\label{remark1}
The derivation of the continuum limit for the static domain case can be carried out in a similar way by assuming $\mathcal{L}_{k}$ to be constant, which implies that $g_{{\bf i}} \equiv 0$ and results in $G \equiv 0$.
\end{remark}
\vspace{-0.3cm}
\section{Set-up of numerical simulations on static domains} 
\label{app:sta}
We let $x\in[0,1]$, $y\in[0,1]$ and $\chi:=0.005$ (\emph{i.e.}, $I=201$). Moreover, we define $\tau :=1\times 10^{-3}$.
\vspace{-0.3cm}
\paragraph{Dynamics of the morphogens} For the dynamics of the morphogens, we consider the parameter setting used in~\cite{maini2012turing}, that is, 
\begin{equation}
D_{u}:=1\times10^{-4}, \quad D_{v}:=4\times10^{-3}, \quad \alpha_{u}:=0.1, \quad \beta:=1, \quad \gamma:=1, \quad \alpha_{v}:=0.9.\label{maini}
\end{equation}
Moreover, we assume the initial distributions to be small perturbations of the homogeneous steady state $(u^*,v^*)\equiv(1,0.9)$, that is,
$$
u^0_{\bf i} = u^* - \rho + 2 \, \rho \, {\bf R} \quad \text{and} \quad v^0_{\bf i} = v^* - \rho + 2 \, \rho \, {\bf R}
$$
where $\rho := 0.001$ and ${\bf R}$ is either a vector for $d=1$ or a matrix for $d=2$ whose components are random numbers drawn from the standard uniform distribution on the interval $(0,1)$, using the built-in {\sc Matlab} function {\sc rand}. These choices of the initial distributions of morphogens are such that 
$$
u^* - \rho \leq u^0_{\bf i} \leq u^* + \rho \quad \text{and} \quad v^* - \rho \leq v^0_{\bf i} \leq v^* + \rho \quad \text{for all } {\bf i},
$$
that is, the parameter $\rho$ determines the level of perturbation from the homogeneous steady state. Since the difference equations~\eqref{eq:discuv} governing the dynamics of the morphogens are independent from the dynamics of the cells, such equations are solved first for all time-steps and the solutions obtained are then used to evaluate both the probabilities of cell movement [\emph{cf.} definitions~\eqref{e:diffusionLR}--\eqref{e:Jdownup}] and the probabilities of cell division and death [\emph{cf.} definitions~\eqref{pb}--\eqref{pq}]. The parameter $u_{\rm max}$ in definitions~\eqref{e:Jleftright} and~\eqref{e:Jdownup} is defined as $\displaystyle{\max_{k, {\bf i}} u_{{\bf i}}^k}$.
\vspace{-0.3cm}
\paragraph{Computational implementation of the rules underlying the dynamics of the cells} At each time-step, each cell undergoes a three-phase process: Phase 1) undirected, random movement according to the probabilities described in the definitions~\eqref{e:diffusionLR} and~\eqref{e:diffusionDU}; Phase 2) chemotaxis according to the probabilities described via the definitions~\eqref{e:Jleftright} and~\eqref{e:Jdownup}; Phase 3) division and death according to the probabilities defined via equations~\eqref{pb}--\eqref{pq}. For each cell, during each phase, a random number is drawn from the standard uniform distribution on the interval $(0,1)$ using the built-in {\sc Matlab} function {\sc rand}. It is then evaluated whether this number is lower than the probability of the event occurring and if so the event occurs.
 \vspace{-0.3cm}
\paragraph{Dynamics of the cells} Unless stated otherwise, we assume the initial cell distributions to be homogeneous with
$$
n^0_i \equiv 10^{4} \; \text{ when } d=1 \quad \text{and} \quad n^0_{\bf i} \equiv 4\times10^{5} \; \text{ when } d=2.
$$ 

In the case where chemically-controlled cell proliferation occurs and there is no chemotaxis, unless stated otherwise, we use the following parameter values when $d=1$
\begin{equation}
\theta:=0.05, \quad \eta:=0,\quad \alpha_{n}:=5, \quad \beta_{n}:=1, \quad n_{\rm max}:=2\times10^{4}.\label{prolifpar1D} \nonumber
\end{equation}
and the following ones when $d=2$
\begin{equation}
\theta:=0.005, \quad \eta:=0,\quad \alpha_{n}:=5, \quad \beta_{n}:=0.1, \quad n_{\rm max}:=8\times10^5.\label{prolifpar2D} \nonumber
\end{equation}
The results shown in Figures~\ref{fig1Dproliferation_lowcell} and \ref{fig2Dproliferation_lowcell} refer to the same settings with the modification that when $d=1$
$$
n^0_i \equiv 4\times10^{3} \quad \text{and} \quad n_{\rm max}:=1.5\times10^{3} 
$$ 
and when $d=2$
$$
n^0_{\bf i} \equiv 2\times10^{5} \quad \text{and} \quad n_{\rm max}:=8\times10^{4}.
$$ 

In the case where cells undergo chemotaxis and cell proliferation is not chemically-controlled, unless stated otherwise, we use the following parameter values when $d=1$
\begin{equation}
\theta:=0.05, \quad \eta:=1,\quad \alpha_{n}:=0.1, \quad \beta_{n}:=0.055, \quad n_{\rm max}:=2\times10^{4}.\label{chemopar1D} \nonumber
\end{equation}
and the following ones when $d=2$
\begin{equation}
\theta:=0.005, \quad \eta=1,\quad \alpha_{n}:=0.1, \quad \beta_{n}:=0.055, \quad n_{\rm max}:=8\times10^5.\label{chemopar2D} \nonumber
\end{equation}
\vspace{-0.3cm}
\paragraph{Numerical solutions of the corresponding continuum models} Numerical solutions of the PDE~\eqref{eq:PDEn} and the system of PDEs~\eqref{eq:PDEuv} subject to zero-flux boundary conditions are computed through standard finite-difference schemes using initial conditions and parameter values that are compatible with those used for the IB model and the system of difference equations~\eqref{eq:discuv}. In particular, the values of the parameters $D_n$ and $C_n$ in the PDE~\eqref{eq:PDEn} are described via the definitions~\eqref{ass:quottozeronum}.
\vspace{-0.3cm}
\section{Set-up of numerical simulations on growing domains} 
\label{app:gro}
We let $x\in[0,1]$, $y\in[0,1]$ and $\chi:=0.005$ ({i.e.}, $I=201$). Moreover, we assume $\tau :=1\times 10^{-3}$ and we define $\mathcal{L}$ according to equation~\eqref{defineL} ({i.e.}, the domain grows linearly over time).
\vspace{-0.3cm}
\paragraph{Dynamics of the morphogens} For the dynamics of the morphogens, we use the parameter setting given by the definitions~\eqref{maini}. Moreover, we define the initial distributions as the numerical equilibrium distributions obtained in the case of static domains. Similarly to the case of static domains, since the difference equations~\eqref{eq:discuvgro} governing the dynamics of the morphogens are independent from the dynamics of the cells, such equations are solved first for all time-steps and the solutions obtained are then used to evaluate both the probabilities of cell movement given by definitions~\eqref{e:diffusionLRgro}--\eqref{e:Jdownupgro} and the probabilities of cell division and death given by equations~\eqref{pbgro}--\eqref{pqgro}. The parameter $u_{\rm max}$ in the definitions~\eqref{e:Jleftrightgro} and~\eqref{e:Jdownupgro} is defined as $\displaystyle{\max_{k, {\bf i}} u_{{\bf i}}^k}$.
\vspace{-0.3cm}
\paragraph{Computational implementation of the rules underlying the dynamics of the cells} Similarly to the case of static domains, at each time-step, each cell undergoes a three-phase process: Phase 1) undirected, random movement according to the probabilities described through the definitions~\eqref{e:diffusionLRgro} and~\eqref{e:diffusionDUgro}; Phase 2) chemotaxis according to the probabilities described through the definitions~\eqref{e:Jleftrightgro} and~\eqref{e:Jdownupgro}; Phase 3) division and death according to the probabilities defined via equations~\eqref{pbgro}--\eqref{pqgro}. For each cell, during each phase, a random number is drawn from the standard uniform distribution on the interval $(0,1)$ using the built-in {\sc Matlab} function {\sc rand}. It is then evaluated whether this number is lower than the probability of the event occurring and if so the event occurs.
\vspace{-0.3cm}
\paragraph{Dynamics of the cells} We assume the initial cell distributions and all parameter values to be the same as those used in the static domain case.
\vspace{-0.3cm}
\paragraph{Numerical solutions of the corresponding continuum models} Numerical solutions of the PDE~\eqref{eq:PDEn_growing} and the system of PDEs~\eqref{eq:PDEuv_growing} subject to zero-flux boundary conditions are computed through standard finite-difference schemes using initial conditions and parameter values that are compatible with those used for the IB model and the system of difference equations~\eqref{eq:discuvgro}. In particular, the values of the parameters $D_n$ and $C_n$ in the PDE~\eqref{eq:PDEn_growing} are described through the definitions~\eqref{ass:quottozeronum}.

\section{Supplementary figures} 

 \begin{figure}[h!]\centering
\includegraphics[width=\textwidth]{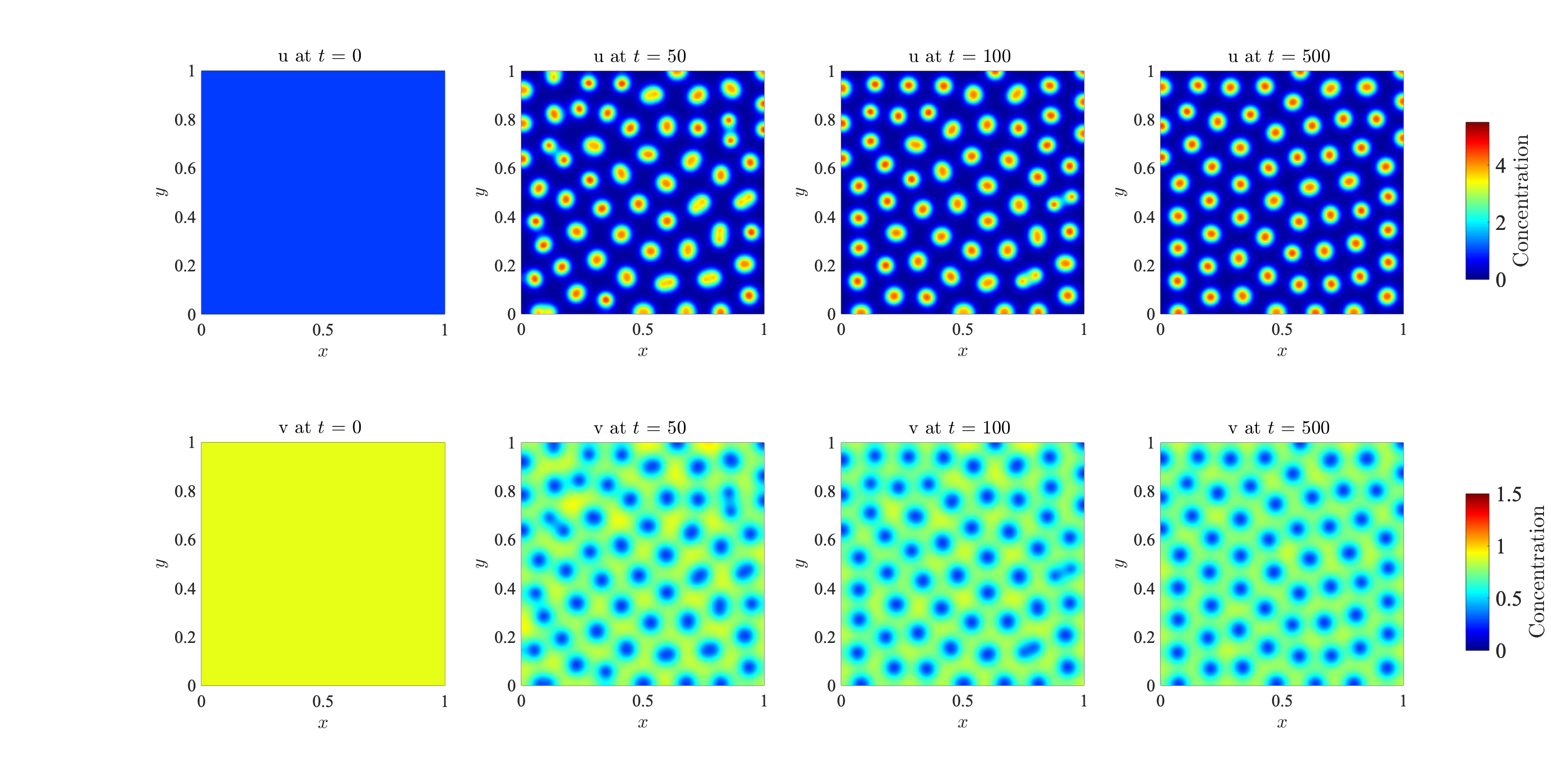}
\caption{{Dynamics of the morphogens on a two-dimensional static domain.} Plots of the concentration of activator $u(t,{\bf x})$ (top row) and the concentration of inhibitor $v(t,{\bf x})$ (bottom row) at four consecutive time instants, obtained by solving numerically the system of PDEs~\eqref{eq:PDEuv} for $d=2$ complemented with the definitions~\eqref{schnackenbergA} and subject to zero-flux boundary conditions. A complete description of the set-up of numerical simulations is given in Appendix~\ref{app:sta}.\label{fig2Dprepattern}
}
\end{figure}
 \begin{figure}[H]\centering
\includegraphics[width=\textwidth]{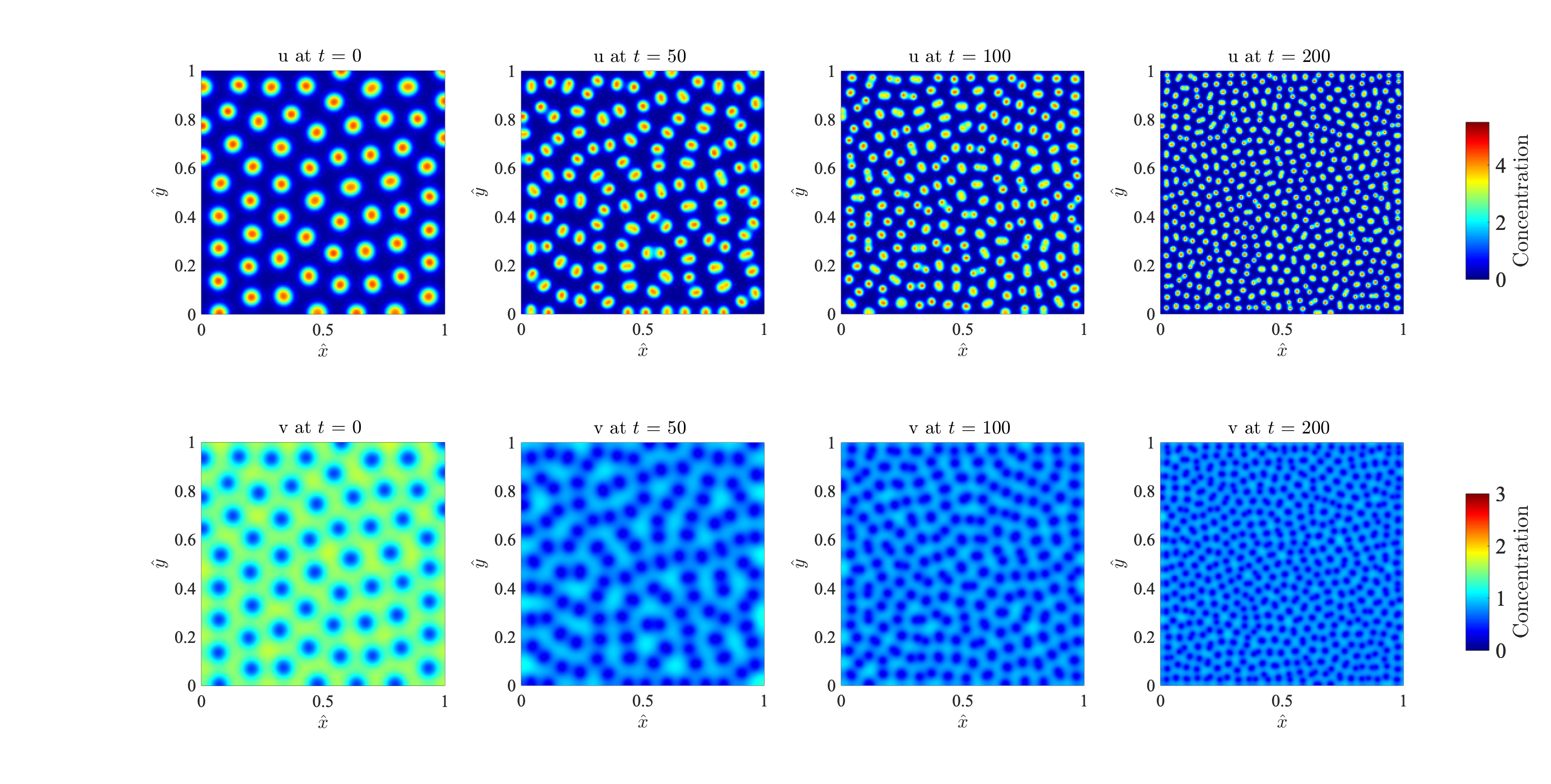}
\caption{{\bf Dynamics of the morphogens on a two-dimensional uniformly growing domain.} Plots of the concentration of activator $u(t,\hat {\bf x})$ (top row) and the concentration of inhibitor $v(t,\hat {\bf x})$ (bottom row) at four consecutive time instants, obtained by solving numerically the system of PDEs~\eqref{eq:PDEuv_growing} for $d=2$, subject to zero-flux boundary conditions, complemented with the definitions~\eqref{schnackenbergA}, equation~\eqref{def:unifdgcont} and equation~\eqref{defineL}. A complete description of the set-up of numerical simulations is given in Appendix~\ref{app:sta}.\label{fig2Dprepatternbothgrowing}
}
\end{figure}

 \begin{figure}[H]\centering
\includegraphics[width=\textwidth]{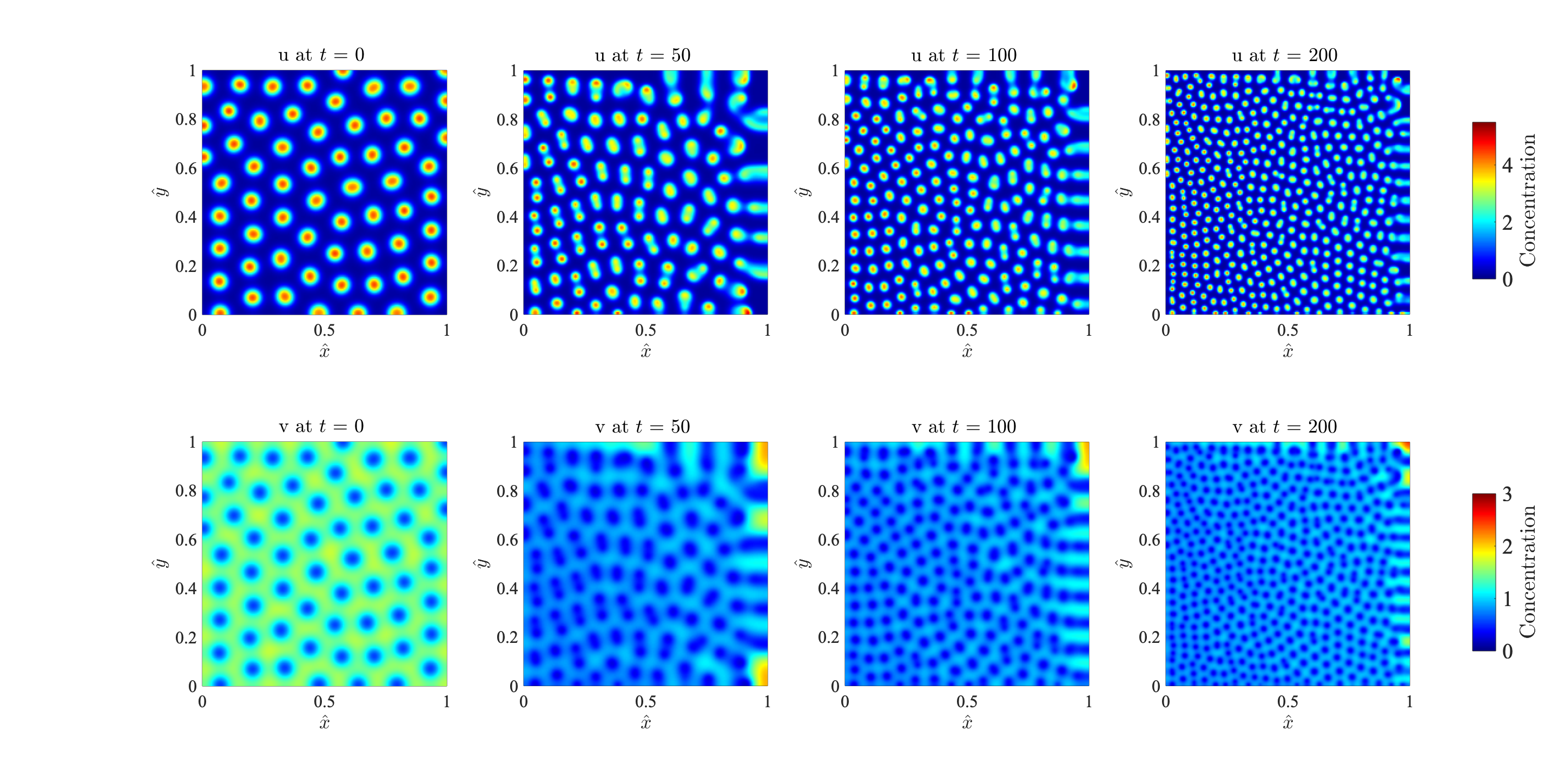}

\caption{{\bf Dynamics of the morphogens on a two-dimensional apically growing domain.} Plots of the concentration of activator $u(t,{\bf x})$ (top row) and the concentration of inhibitor $v(t,{\bf x})$ (bottom row) at four consecutive time instants, obtained by solving numerically the system of PDEs~\eqref{eq:PDEuv_growing} for $d=2$, subject to zero-flux boundary conditions, complemented with the definitions~\eqref{schnackenbergA}, equation~\eqref{def:apicalcont} and equation~\eqref{defineL}. A complete description of the set-up of numerical simulations is given in Appendix~\ref{app:sta}.\label{fig2DprepatternbothgrowingTip}
}
\end{figure}

\end{document}